\documentclass[twocolumn,english,prl,twocolumn,floatfix,footinbib,superscriptaddress,showpacs, showkeys,notitlepage]{revtex4-1}
\usepackage[latin9]{inputenc}
\usepackage{amsmath}
\usepackage{amssymb}
\usepackage{graphicx}
\usepackage{esint}

\makeatletter
\@ifundefined{textcolor}{}
{%
 \definecolor{BLACK}{gray}{0}
 \definecolor{WHITE}{gray}{1}
 \definecolor{RED}{rgb}{1,0,0}
 \definecolor{GREEN}{rgb}{0,1,0}
 \definecolor{BLUE}{rgb}{0,0,1}
 \definecolor{CYAN}{cmyk}{1,0,0,0}
 \definecolor{MAGENTA}{cmyk}{0,1,0,0}
 \definecolor{YELLOW}{cmyk}{0,0,1,0}
}

\global\long\def\ket#1{\left| #1\right\rangle }

\global\long\def\bra#1{\left\langle #1 \right|}

\global\long\def\kket#1{\left\Vert #1\right\rangle }

\global\long\def\bbra#1{\left\langle #1\right\Vert }

\global\long\def\braket#1#2{\left\langle #1\right. \left| #2 \right\rangle }

\global\long\def\bbrakket#1#2{\left\langle #1\right. \left\Vert #2\right\rangle }

\global\long\def\av#1{\left\langle #1 \right\rangle }

\global\long\def\tr{\text{Tr}}

\global\long\def\im{\text{Im}}

\global\long\def\re{\text{Re}}

\newcommand{\be}{\begin{eqnarray}}\newcommand{\ee}{\end{eqnarray}}\def\beq{\begin{equation}}\def\eeq{\end{equation}}

\usepackage{babel}

\usepackage{babel}

\makeatother

\usepackage{babel}
\begin{document}

\title{Mott insulator breakdown through pattern formation}

\author{Pedro Ribeiro}

\affiliation{Russian Quantum Center, Novaya street 100 A, Skolkovo, Moscow area,
143025 Russia}

\email{ribeiro.pedro@gmail.com}

\author{Andrey E. Antipov}

\affiliation{Department of Physics University of Michigan, Randall Laboratory,
450 Church Street, Ann Arbor, MI 48109-1040}

\author{Alexey N. Rubtsov }

\affiliation{Russian Quantum Center, Novaya street 100 A, Skolkovo, Moscow area,
143025 Russia}
\begin{abstract}
We study the breakdown of a Mott insulator with the thermodynamic
imbalance induced by an applied bias voltage. By analyzing the instabilities
of the magnetic susceptibility, we describe a rich non-equilibrium
phase diagram, obtained for different applied voltages, that exhibits
phases with a spatially patterned charge gap. For a finite voltage,
smaller than the value of the equilibrium Mott gap, the formation
of patterns coincides with the emergence of mid-gap states contributing
to a finite steady-state conductance. We discuss the experimental
implications of this new scenario of Mott breakdown. 
\end{abstract}


\maketitle
\begin{minipage}[t]{1\columnwidth}%
\global\long\def\ket#1{\left| #1\right\rangle }

\global\long\def\bra#1{\left\langle #1 \right|}

\global\long\def\kket#1{\left\Vert #1\right\rangle }

\global\long\def\bbra#1{\left\langle #1\right\Vert }

\global\long\def\braket#1#2{\left\langle #1\right. \left| #2 \right\rangle }

\global\long\def\bbrakket#1#2{\left\langle #1\right. \left\Vert #2\right\rangle }

\global\long\def\av#1{\left\langle #1 \right\rangle }

\global\long\def\tr{\text{tr}}

\global\long\def\Tr{\text{Tr}}

\global\long\def\pd{\partial}

\global\long\def\im{\text{Im}}

\global\long\def\re{\text{Re}}

\global\long\def\sgn{\text{sgn}}

\global\long\def\Det{\text{Det}}

\global\long\def\abs#1{\left|#1\right|}

\global\long\def\up{\uparrow}

\global\long\def\down{\downarrow}

\global\long\def\k{\mathbf{k}}

\global\long\def\wks{\mathbf{\omega k}\sigma}

\global\long\def\vc#1{\mathbf{#1}}

\global\long\def\bs#1{\boldsymbol{#1}}
\end{minipage}Pattern formation, also known as self-organization, refers to the
occurrence of spatial-structured steady-states in non-linear systems
under out of equilibrium external conditions \cite{Cross1993}. A
textbook illustration is the Rayleigh\textendash Bénard convection,
but examples are found ubiquitously in physical, chemical as well
as in biological systems \cite{Nicolis1977,Ball}. 

In semiconductors, pattern formation is a hallmark of the voltage-driven
non-equilibrium phase transition from insulating to the metallic state
\cite{Scholl1987}, where moving patterns arise near phase boundaries
that contribute to the finite conductivity of the system. A seminal
experiment, revealing pattern formation in strongly correlated systems
\cite{Kumai1999} reported a current-induced pattern formation in
a quasi-one dimensional organic charge-transfer complex, on the verge
of Mott breakdown. A non-linear I-V characteristic was reported in
a low-resistance state characterized by a striped charge pattern,
before the switching to metallic regime. Recently, experimental results
for spinor Bose-Einstein condensates \cite{Kronjager2010} and, theoretical
studies of polariton condensates \cite{Borgh2010,Berloff2013} also
reported patterned phases.

Non-equilbrium dynamics of strongly correlated quantum many-body systems
have been recently receiving an increased attention due to a rich
interplay between electronic kinetics, interaction and non-equilibrium
conditions. Major experimental progress was driven forward by a tight
control of the dynamics in cold atomic setups \cite{Bloch2008,Strohmaier2010}
and pump-probe experiments \cite{Cavalleri2001,Novelli2014}. On the
theory side, progress been done in understanding thermalization and
dissipation \cite{Rigol2008b,Srednicki1994,Deutsch1991}, universal
aspects of non-equilibrium phase transitions \cite{Diehl2008,Diehl2010,Sieberer2013,Mitra2006,Mitra2008a,Takei2010,Chung2009,Kirchner2009a,Ribeiro2013b}
and the development of involved computational methods \cite{Werner2009,Schiro2009,Gull2010,Gull2011,Cohen2013}
and techniques \cite{Aoki2014,Schiro2010,Schiro2011}. In particular,
the study of out-of-equilibrium properties of the Hubbard model has
been an active research area \cite{Eckstein2010,Aron2012a,Arrigoni2013,Aoki2014}.
Interesting dynamical transitions between small and large interaction
quenches where shown to occur at half-filling \cite{Moeckel2008,Eckstein2009,Schiro2011,Schiro2010,Enss2012}.
Transport properties at finite temperature \cite{Karrasch2014} and
in the presence of Markovian dissipation \cite{Prosen2012,Prosen2014}
have been investigated. 

A key problem is the understanding of the transition from a Mott insulator
to a current-carrying state upon applied an increasing voltage bias
to coupled external leads. The generated electro-chemical gradients
induce two effects of rather different nature: (i) a thermodynamic-imbalance
depending on the distribution functions of the leads and (ii) the
coupling of the charged particles to the electric field created by
the voltage drop. 

The breakdown of a Mott insulator induced by effect (ii) recently
received important contributions. Using Peierls substitution argument,
(ii) can be studied on a system with periodic boundary conditions
pierced by a linear-in-time magnetic flux, eliminating the need of
explicitly treating the reservoirs and making it amenable to be tackled
by Lanczos \cite{Oka2003}, DMRG \cite{Oka2005} and DMFT \cite{Eckstein2010b,Eckstein2011a,Aron2012a}
methods. These studies revealed a qualitative scenario \cite{Oka2003}
interpreted as the many-body analog of the Landau-Zener (LZ) mechanism
observed in band insulators. The LZ energy scale sets a threshold
$V_{\text{th}}\sim\Delta^{2}L/W$, with $\Delta$ being the Mott gap,
$L$ -- the system's linear size and $W$ -- the bandwidth, above
which a field-induced metallic phase sets in. Zener\textquoteright s
formula yields $V_{\text{th}}/L\gg\Delta$ overestimating experimental
values of threshold fields. 

The combined effect of (i) and (ii) have also been recently addressed
\cite{Sugimoto2008,Heidrich-Meisner2010,Tanaka2011}. As (i) requires
the explicit treatment of the reservoirs, non-equilibrium Green's
functions approaches were employed. (ii) was treated within the Hartree
approximation with a fixed antiferromagnetic order, precluding any
pattern formation. The results are compatible with a current-voltage
characteristics of the form $J\simeq Ve^{-V_{\text{th}}/V}$. A thorough
study \cite{Tanaka2011}, carried out at $T=0$ in the presence of
long-range Coulomb interactions, pointed out that the dominant effect
depends on the ratio between the correlation length in the insulating
phase $\xi$ and the size of the insulating region $L$. For $\xi/L\gg1$,
(i) leads to $V_{\text{th}}\sim\Delta$; for $\xi/L\ll1$ (ii) dominates
and the LZ scenario is recovered. 

In this letter, we address out-of-equilibrium properties of Hubbard
chain due to thermodynamic-imbalance (i). We describe the appearance
of mobile carriers that contribute to the screening of the field.
The leads provide, at the same time, the non-equilibrium conditions
and an intrinsically non-Markovian \cite{RibeiroVieira2014} dissipative
environment. We compute the instabilities of the system to spatially
modulated spin patterns and identify a rich set of candidate phases,
among which examples of pattern formation, analyzing their properties
in the strong nonlinear regime. We put forward a scenario of the Mott
breakdown through the emergence of conducting mid-gap states coinciding
with the appearance of patterns for $V_{\text{th}}\lesssim\Delta$.
Our results are of relevance to pattern formation in quasi-one dimensional
organic compounds \cite{Kumai1999}.

\begin{figure}
\begin{centering}
\includegraphics[width=1\columnwidth]{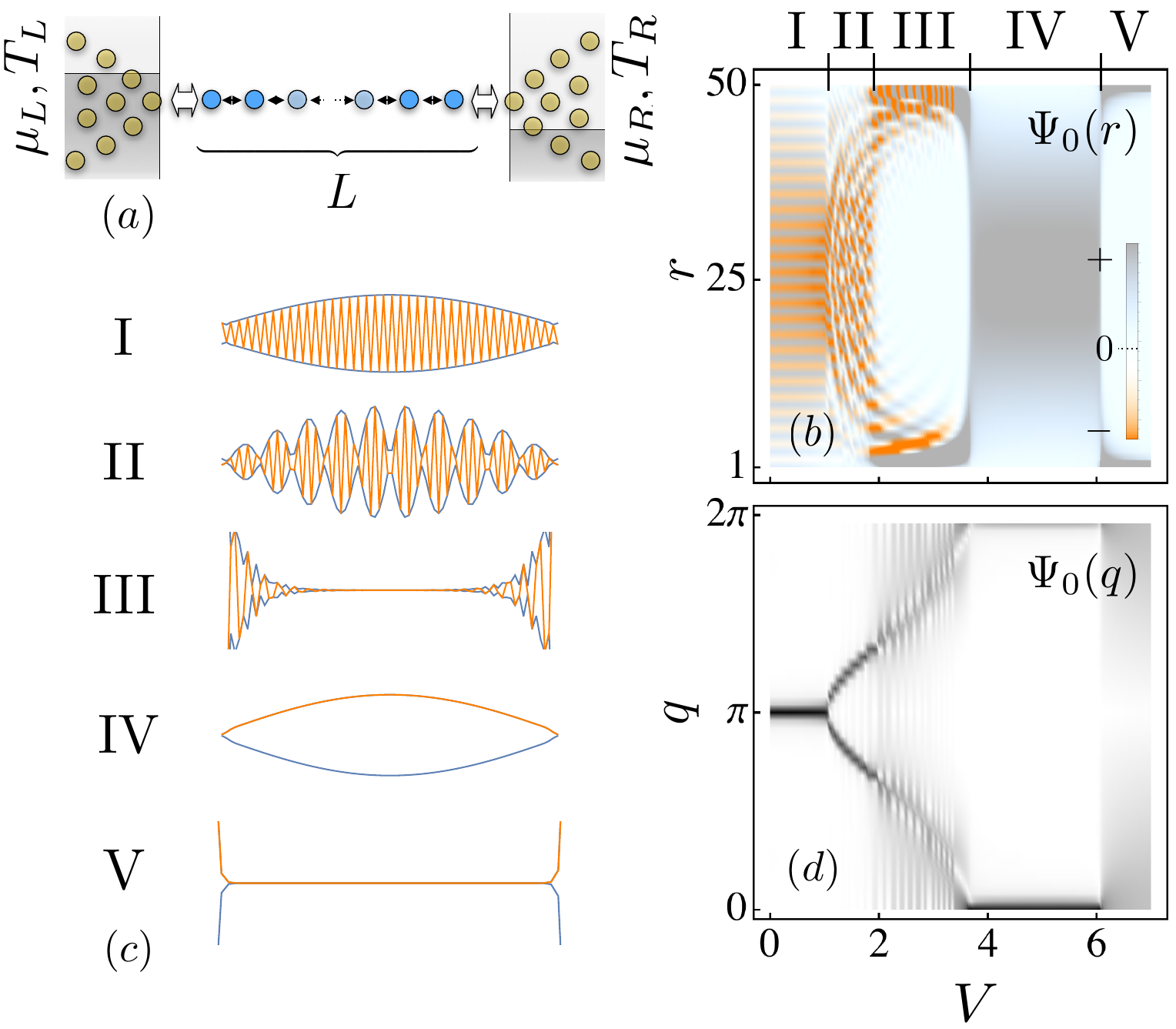}
\par\end{centering}

\protect\caption{\label{fig:real_and_k_space}(a) Schematic view of the physical setup.
(b) Density plot of the most unstable mode $\Psi_{0}\left(r\right)$plotted
as a function of the bias $V$ for $\Gamma=0.25$, $T=0.25$ and $L=50$.
The phase labels I,...,V point to qualitatively different behavior
of $\Psi_{0}\left(r\right)$. (c) Typical spatial dependence of $\Psi_{0}\left(r\right)$
in each phase (orange line), plotted for $L=80$. The blue line depicts
the envelope function. (c) Density plot of the Fourier transform $\Psi_{0}\left(q\right)$
of $\Psi_{0}\left(r\right)$ as a function of $q$ computed for $L=50$. }
\end{figure}

We consider the interacting system $\text{S}$, in Fig. \ref{fig:real_and_k_space}-(a),
consisting of a chain coupled to metallic reservoirs. The Hamiltonian
can be decomposed as $H=H_{\text{S}}+H_{\text{\ensuremath{\pd}}\text{S}}+H_{\bar{\text{S}}}$,
where 
\begin{eqnarray}
H_{\text{S}} & = & -t\sum_{\langle\bs r,\bs r'\rangle,s}c_{\bs rs}^{\dagger}c_{\bs{r'}s}+\frac{U}{2}\sum_{\bs r}\left(n_{\bs r}-1\right)^{2}
\end{eqnarray}
is the Hamiltonian of the system consisting of a fermionic Hubbard
chain, with $s$ labeling spin degrees of freedom and $n_{\bs r}=\sum_{\sigma}c_{\bs rs}^{\dagger}c_{\bs rs}$.
The hopping matrix element between nearest neighbor sites, $t=1$,
is taken to be the energy unit. $H_{\bar{\text{S}}}=\sum_{\alpha,s,l}d_{l\alpha s}^{\dagger}\epsilon_{l,\alpha}d_{l\alpha s}$
is the Hamiltonian of the reservoirs, with $l=L,R$ labeling the reservoir
and $\alpha$ -- the reservoir's single-particle modes. The density
of states of the leads is taken to be the one of a wide band metallic
lead, i.e. a constant $\rho$, within all the considered energy scales
for both leads. The system-reservoirs coupling is described by the
hopping term $H_{\pd\text{S}}=\sum_{\alpha,s,l}v\, d_{l\alpha s}^{\dagger}c_{\bs r_{l},s}+\text{h.c.}$,
where $\bs r_{L,R}$ are the sites at the extremities of the chain
and $v$ is the hopping amplitude taken to be spin independent. We
consider reservoirs at temperature $T$ that are characterized by
the same hybridization $\Gamma=\pi v^{2}\rho$ for simplicity. 

We employ a non-equilibrium mean-field approach, that while providing
only a qualitative description of the 1d model, allows to probe instabilities
of the system towards the formation of gapped phases. The procedure
to obtain the mean-field equations and the magnetic susceptibility
is standard and is given in the SI for completeness. Here we outline
the main steps. Working on the Keldysh contour we use the identity
$\frac{U}{2}\sum_{\bs r}\left(n_{\bs r}-1\right)^{2}=-\frac{3}{4}U\left(\bs S_{\bs r}.\bs S_{\bs r}-1\right)$,
with $\bs S_{\bs r}=\frac{1}{2}c_{\bs r,s}^{\dagger}\bs{\sigma}_{ss'}c_{\bs r,s'}$,
and insert a 3-component time dependent order-parameter $\bs{\phi}(t)$
to decouple the interaction term in the spin-density wave channel
$\frac{3}{4}U\bs S_{\bs r}.\bs S_{\bs r}\to\bs S_{\bs r}.\bs{\phi}_{\bs r}+\frac{1}{3U}\bs{\phi}_{\bs r}.\bs{\phi}_{\bs r}$.
Assuming a wide-band limit, we then integrate out the non-interacting
reservoirs introducing a local self-energy contribution for the interacting
$c$ electrons with non-zero components (see SI-sec.\ref{sub:Properties-of-the}):
$\Sigma_{\bs r=\bs r_{l},\bs r'=\bs r_{l}}^{R/A}\left(t,t'\right)\simeq\mp i\Gamma\delta\left(t-t'\right)$,
$\Sigma_{\bs r=\bs r_{l},\bs r'=\bs r_{l}}^{K}\left(t,t'\right)\simeq-2i\Gamma\int\frac{d\varepsilon}{2\pi}\tanh\left[\frac{\beta_{l}}{2}\left(\varepsilon-\mu_{l}\right)\right]e^{-i\varepsilon(t-t')}$.
Integrating out the $c$ electrons, we arrive to the action for the
order-parameter $\bs{\phi}(t)$ alone. We use the Keldysh rotation
of the time-dependent order parameter to the quantum and classical
components ($\bs{\phi}_{c,\bs r}$,$\bs{\phi}_{q,\bs r}$) and by
varying the action with respect to these fields we obtain their mean
field values:

\begin{equation}
\begin{aligned}\bs{\phi}_{c,\bs r}\left(t\right) & =-i\frac{3U}{4}\frac{1}{\sqrt{2}}\tr\left[G_{\bs r\bs r}^{K}\left(t,t\right)\bs{\sigma}\right]\\
\bs{\phi}_{q,\bs r}(t) & =0,
\end{aligned}
\label{eq:mean-field}
\end{equation}
where $G_{\bs r\bs r}^{K}\left(t,t\right)$ is the Keldysh component
of the local $c$-electron Green's function. We focus on the steady
state regime $\bs{\phi}_{c,\bs r}\left(t\right)=\bs{\phi}_{c,\bs r}$.
At the mean-field level, the excitation spectrum is given by the non-hermitian
mean-field operator 
\begin{multline}
K=-t\sum_{\langle\bs r,\bs r'\rangle,s}c_{\bs rs}^{\dagger}c_{\bs{r'}s}-i\Gamma\sum_{l,s}c_{\bs{r_{l}}s}^{\dagger}c_{\bs{r_{l}}s}-\\
-\frac{1}{\sqrt{2}}\sum_{\bs rss'}\left(\bs{\sigma}_{ss'}.\bs{\phi}_{c,\bs r}\left(t\right)\right)c_{\bs rs}^{\dagger}c_{\bs rs'}.
\end{multline}
The retarded Green's function is obtained as a function of the left-
($\bra{\tilde{\alpha}}$) and right- ($\ket{\alpha}$) eigenvectors
of $K$ with complex eigenvalues $\lambda_{\alpha}$ ($\im\lambda_{\alpha}<0$):
$G^{R}\left(\omega\right)=\sum_{\alpha}\ket{\alpha}\left(\omega-\lambda_{\alpha}\right)^{-1}\bra{\tilde{\alpha}}$.
The Keldysh component, derived in detailed in the SI, is obtained
in a similar way.

Fluctuations around the mean-field further provide a stability analysis
for the saddle-point solutions. In order to investigate the possible
steady-states that can be realized under non-equilibrium conditions
we compute the spin susceptibility $\chi$ in the disordered state
($\bs{\phi}_{c,\bs r}=0$) and analyze the first unstable modes arising
upon increasing $U$. The retarded spin susceptibility $\chi_{ii';\bs r\bs r'}^{R}\left(t,t'\right)=-i\Theta\left(t-t'\right)\av{\Bigl\{ S_{\bs r}^{i}(t),S_{\bs r'}^{i'}(t')\Bigr\}}$
(with $i,i'=x,y,z$) is given by the RPA-type expression and in the
steady state reads
\begin{eqnarray}
\left[\chi_{ii'}^{R}\left(\omega\right)\right]_{\bs r\bs r'}^{-1} & = & \delta_{ii'}\left[-\frac{2}{3U}\delta_{\bs r\bs r'}-\Xi_{\bs r\bs r'}^{R}\left(\omega\right)\right],\label{eq:inverse_sus}
\end{eqnarray}
where $\Xi_{rr'}^{R}\left(t,t'\right)=-i\frac{1}{2}\tr[G_{\bs r'\bs r}^{A}\left(t',t\right)G_{\bs r\bs r'}^{K}\left(t,t'\right)+G_{\bs r'\bs r}^{K}\left(t',t\right)G_{\bs r\bs r'}^{R}\left(t,t'\right)]$
is the bare bubble diagram computed at $\bs{\phi}_{c,\bs r}=0$ and
$G_{\bs r\bs{r'}}^{R/A}\left(t,t'\right)$ are the spatially resolved
retarded/advanced components of the Green's function of the $c$-electrons.

Upon increasing $U$, the eigenvalues of $\chi^{R}\left(\omega\right)$
as a functions of $\omega$, may develop poles in the upper-half of
the complex plane. When this occurs, small perturbations in the direction
of the corresponding eigenmode of $\chi^{R}\left(\omega\right)$ grow
exponentially in time until anharmonic mode-coupling terms start to
be relevant. This process signals an instability of the system. The
new stable phase, arising for $U>U_{c}$, is expected to develop the
spatial structure of the lowest eigen-mode of $\chi^{R}\left(\omega\right)$,
at least for $U$ sufficiently close to $U_{c}$. In the following
we assume that unstable modes first occur for steady-state solutions
i.e. at $\omega=0$. The unstable mode corresponds to the most negative
eigenvalue $\lambda_{0}^{\Xi}$ of $\bs{\Xi}^{R}\left(\omega=0\right)$
and its spatial configuration is given by the corresponding eigenvector
$\Psi_{0}\left(r\right)$. 
\begin{figure}
\begin{centering}
\includegraphics[width=1\columnwidth]{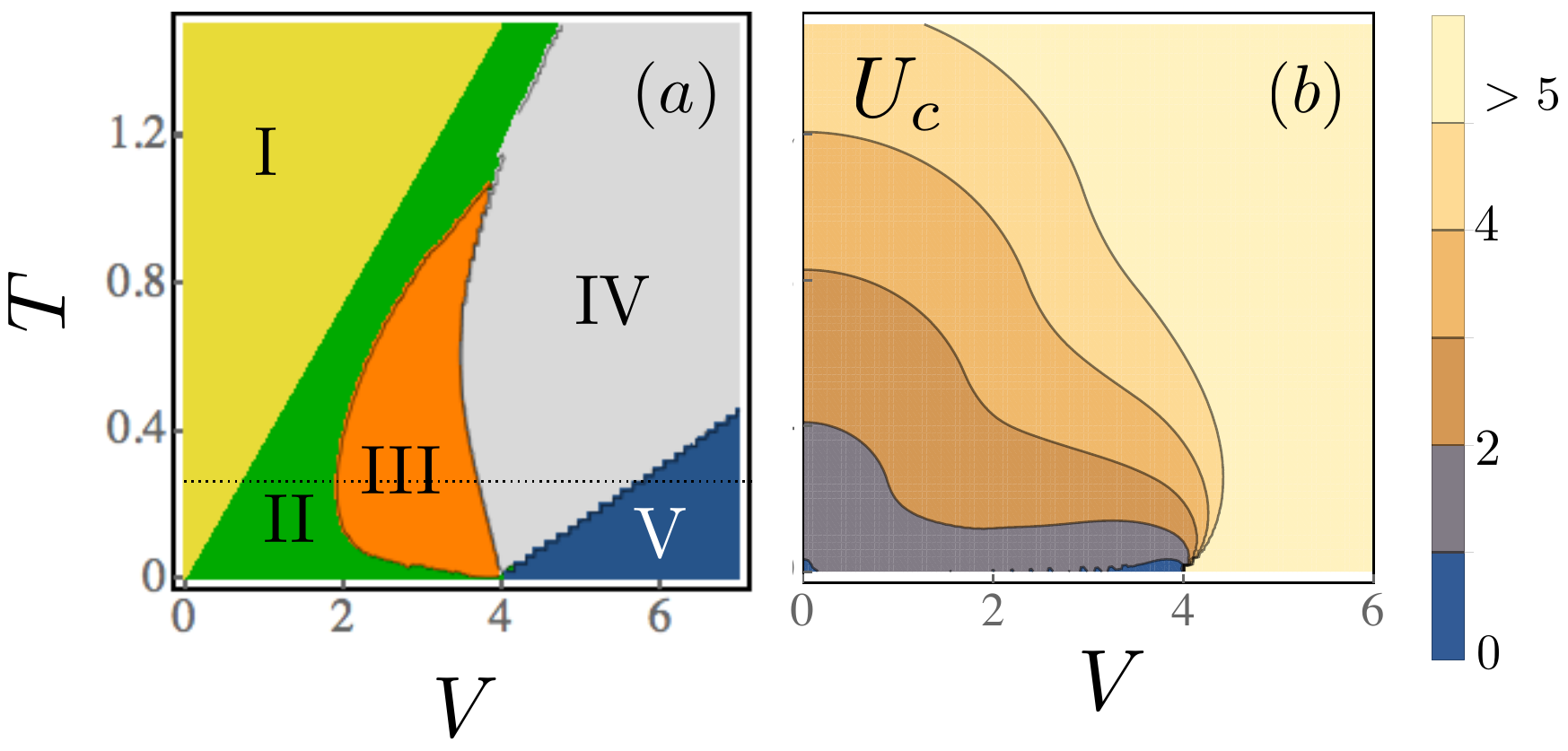}
\par\end{centering}

\protect\caption{\label{fig:phase-d}(a) Phase diagram as a function of $V$ and $T$
computed for $\Gamma=0.25$ and $U=U_{c}\left(T,V\right)$. The dashed
line corresponds to the plots (b) and (c) of Fig.\ref{fig:real_and_k_space}.
(b) Values of $U_{c}$ for which the first instability arises as a
function of $V$ and $T$, for $\Gamma=0.25$ and $L=50$. }
\end{figure}

At equilibrium, and for periodic boundary conditions, $\Psi_{0}\left(r\right)=\frac{1}{\sqrt{L}}e^{iQr}$,
with $Q=\pi$, signals the instability to the antiferromagneticaly
ordered phase. This picture is essentially unchanged in the presence
of open boundary conditions with the order parameter amplitude typically
getting distorted near the boundaries of the system. 

Figs. \ref{fig:real_and_k_space}-(b,c) depict the typical spatial
structure of steady state $\Psi_{0}\left(r\right)$ obtained upon
varying the bias voltage $V$. Five different phases (labeled by I,...,V)
can be observed, corresponding to qualitatively different features
of $\Psi_{0}\left(r\right)$. Fig. \ref{fig:real_and_k_space}-(d)
depicts a contour plot of the Fourier transform $\Psi_{0}\left(q\right)$
of $\Psi_{0}\left(r\right)$ showing that the different phases correspond
to different wave vectors $Q$ for which $\abs{\Psi_{0}\left(Q\right)}$
is maximal. Phase I occurs for low voltages $V<V_{\text{AF}}$ and
$T>0$ and occupies a region where the antiferromagnetic phase corresponds
to the first instability. The order parameter is maximal in the center
of the system. The emergence of patterns is visible in phase II ($V_{\text{AF}}<V<V_{\text{loc }}$),
where the spin-susceptibility instability corresponds to an ordered
state with wave vectors $q=\pm Q$, with $Q$ varying between $\pi$,
for $V=V_{\text{AF}}$, and $Q\leq0$, for $V=V_{\text{loc}}$. Phase
III ($V_{\text{loc}}<V<V_{\text{F}}$) corresponds to a modulated
phase, with $Q\neq0,\pi$, exponentially localized near the leads.
Phase IV ($V_{\text{F}}<V<V_{0}$) is a ferromagnetic phase with an
envelope function that is maximal at the center of the system. Finally,
phase V corresponds to an essentially disordered phase ($\phi=0$)
with the order parameter amplitude being localized in the first few
sites near the leads. 

Fig. \ref{fig:phase-d}-(a) shows the phase diagram in the $V-T$
plane for $\Gamma=0.25$ near $U=U_{c}(T,V)$ for which the first
instability arises. At $T=0$ the anti-ferromagnetism of phase I is
unstable under any finite bias voltage giving place to the modulated
phase II. Moreover, at zero temperature no ferromagnetic phase is
present yielding a direct transition form II to the disordered phase
V. The localized modulated phase III is present only for intermediate
temperatures. For sufficiently high temperatures, within the range
of temperatures and voltages studied, only phase I, II and IV are
observed. The critical value of $U$, given by $U_{c}=-2/(3\lambda_{0}^{\Xi})$
after Eq.(\ref{eq:inverse_sus}), is plotted in Fig.(\ref{fig:phase-d})-(b).
for a system with $L=50$. For low temperature, this quantity is subjected
to strong finite size corrections. Care must be taken extrapolating
to the thermodynamic limit, nonetheless we verify that for $T\to0$
and $L\to\infty$ one has $U_{c}\to0$. 

\begin{figure}
\begin{centering}
\includegraphics[width=1\columnwidth]{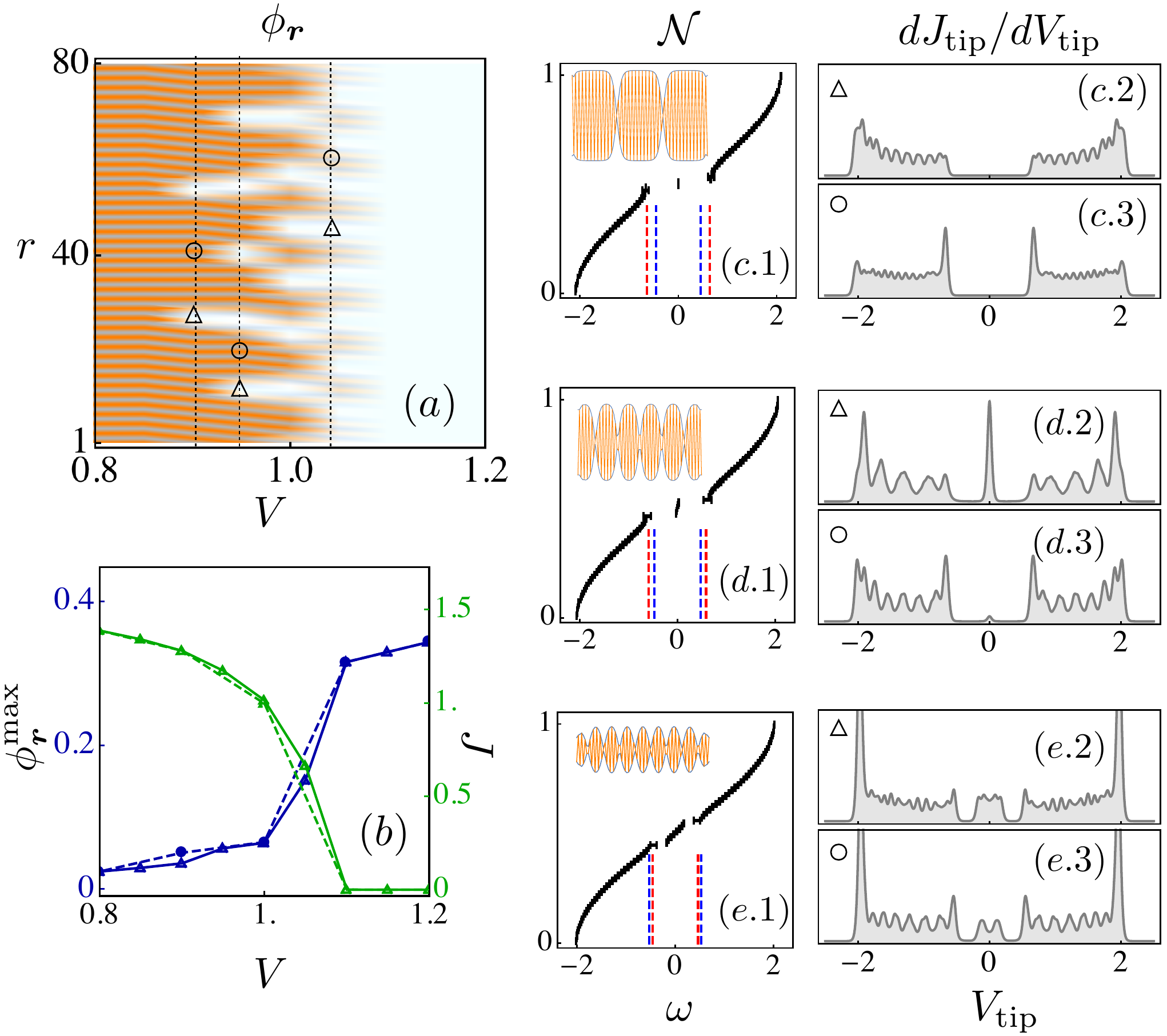}
\par\end{centering}

\protect\caption{\label{fig:order_p} Properties for $U>U_{c}$ obtained for $\Gamma=0.25$,
$T=0.25$, $U=3.8$ corresponding to an equilibrium ($V=0$) Mott
gap of $\Delta=2\protect\abs{\phi}\simeq3.2$. (a) Density plot of
$\Phi\left(r\right)$ plotted as a function of $V$ for $L=80$. The
lines and markers label the specific values of Figs. (c-e). (b) Maximum
value of the order parameter $\phi_{\text{Max}}=\text{max}_{r}\protect\abs{\phi\left(r\right)}$
(green) and particle current thought the chain $J$ (blue) as a function
of $V$ for $L=80$ (open triangles) and $L=120$ (full circles).
(c.1) Integrated density of states $\mathcal{N}\left(\omega\right)=\sum_{\alpha}\Theta\left(\omega-\protect\re\lambda_{\alpha}\right)$
for $V=0.9$ and $L=80$, the thickness of the black line is given
by $\protect\im\lambda_{\alpha}$. The red-dashed lines correspond
to $\omega=\pm\Phi_{\text{Max}}$ and the blue-dashed lines to $\omega=\pm V/2$.
The inset depicts the spatial dependence of $\phi\left(r\right)$.
(c.2-3) Differential conductance $dJ_{\text{tip}}/dV_{\text{tip }}$
obtained by an STM tip, computed for $T_{\text{tip}}=0.02$, placed
at position $\protect\bs r$, for $\protect\bs r=27$ (c.2) $\protect\bs r=41$
(c.3), corresponding to a minimum and a maximum of the order parameter
amplitude. (d.1-3) Same as (c.1-3) for $V=0.95$, $\protect\bs r=12$
and $\protect\bs r=19$. (e.1-3) Same as (c.1-3) for $V=1.05$, $\protect\bs r=45$
and $\protect\bs r=59$. }
\end{figure}

In order to verify the existence of well-defined patters at $U>U_{c}$
and describe their spatial structure, the linear response RPA-type
description is insufficient, as non-linear terms in Eq.(\ref{eq:mean-field})
start to play an important role and have to be taken into account.
In this regime, the mean-field solution for the order parameter $\bs{\phi}$
is obtained solving the self-consistent relation in Eq.(\ref{eq:mean-field}).
The procedure is done iteratively allowing only for collinear magnetized
states, i.e. $\av{\bs S_{\bs r}}\propto\hat{\bs e}_{z}$. Fig.\ref{fig:order_p}-(a)
shows the spatial structure of $\phi\left(r\right)$ obtained in this
way. The considered value of $U=3.8$ corresponds to an equilibrium
($V=0$) Mott gap of $\Delta=2\abs{\phi}\simeq3.2$. Out of equilibrium,
phases III-V are absent and the range of values of $V$ for which
phase II arises is reduced with respect to the diagram of Fig. \ref{fig:phase-d}-(a).
Nevertheless, a modulated solution can be found deep into the non-linear
regime. Fig.\ref{fig:phase-d}-(b) depicts the maximum value of the
order parameter amplitude $\phi_{\text{Max}}$ showing that phase
II transits directly to the disordered phase $\phi=0$ upon increasing
$V$. 

Fig. \ref{fig:phase-d}-(b) shows also the values of the particle
current through the system. A relatively low current in phase I is
followed by a quick rise of current during phase II and a linear I-V
characteristics in the disordered phase. Figs. \ref{fig:phase-d}-(c-e.1)
show the integrated steady state density of states in phase II. One
observes that upon increasing $V$ a new band of conducting states
arises, corresponding to single particle-energies $-V/2<\re\lambda_{\alpha}<V/2$.
The appearance of such states is responsible for the current increase
in phase II. This phase ceases to exist when $V$ becomes of the order
of the of the inter-band gap, roughly given by $\phi_{\text{Max}}$,
corresponding a complete filling of the gap by conducting states.
The I-V characteristics can thus be used to discriminate between different
behaviors. 

To further characterize these states we monitor the differential conductivity
that is measured by an STM tip placed over site $\bs r$. Assuming
a wide-band metallic tip with constant DOS $\rho_{\text{tip}}$, weakly
coupled to the chain at position $\bs r$ by an hopping amplitude
$t_{\text{tip}}$, one obtains the standard linear-response expression
\begin{eqnarray*}
\frac{dJ_{\text{tip}}}{dV_{\text{tip}}} & \propto & -\int d\omega\frac{\beta_{\text{tip}}/2}{\cosh\left[\beta_{\text{tip}}\left(\omega-V_{\text{tip}}\right)\right]+1}\rho_{\bs r}\left(\omega\right)
\end{eqnarray*}
where $\rho_{\bs r}\left(\omega\right)=\tr\left[G_{\bs r,\bs r}^{R}\left(\omega\right)-G_{\bs r,\bs r}^{A}\left(\omega\right)\right]/\left(-2\pi i\right)$
is the local DOS of the chain at site $\bs r$, $\beta_{\text{tip}}$
and $V_{\text{tip}}$ are respectively the tip's inverse temperature
and chemical potential. Figs. \ref{fig:order_p} (c-e.2-3) show $dJ_{\text{tip}}/dV_{\text{tip}}$
for sites corresponding to minima and maxima of the order parameter
for 3 values of $V$ within phase II. The band of conducting states
is can clearly be seen arising within the gap. The local DOS for $\abs{V_{\text{tip}}}<\phi_{\text{Max}}$
increases or decreases, depending on whether a position corresponding
to a minimum or a maximum of the order parameter amplitude is monitored. 

To summarize, we have described a scenario of the Mott breakdown,
induced by the pattern formation in a correlated electronic system
under strong non-equilibrium conditions imposed by a finite bias voltage.
The development of a conducting phase occurs at voltages, smaller
than the value of the charge gap, and is characterized by the emergence
of the mid-gap states. The thermodynamic imbalance imposed by a finite
applied voltage generates a rich set of behaviors, among which examples
of non-equilibrium spatially-induced patterned phases. Such phases,
well studied in classical systems, and recently predicted in systems
with Markovian dissipation \cite{Borgh2010,Berloff2013}, are here
reported for the fermionic Hubbard model with a non-Markovian environment
and are shown to exist down to zero temperature. The suggested mechanism
can be tested experimentally monitoring current transport across the
system and by STM measurements, spatially resolving the modulated
charge gap. 

Our considerations capture characteristic features of the breakdown
of the organic charge insulator, reported in Ref. \cite{Kumai1999}.
The transition to the conducting state, accompanied by the formation
of alternating carrier rich stripes, is reproduced with a similar
I-V characteristic. Important differences, such as a diffusive electronic
transport and the long-range Coulomb interactions within the Mott
phase, hinder a quantitative prediction of the experimental parameters. 

The present results suggest that, as in the case of classical systems,
patterned phases can be ubiquitous in the presence of interactions
and spatially non-uniform out of equilibrium conditions. In 1d, the
phase transitions obtained at the mean-field level should instead
correspond to crossovers. In the same way, the calculated magnetic
order is likely to correspond to a disordered phase with slow power-law
decaying spin-spin correlation functions with a voltage-dependent
$Q$. The emergent order, seen at the mean-field level, can otherwise
be stabilized by weakly coupling multiple chains. For electronic systems
with higher dimensionality, such as films and bulk compounds, pattern
formation should naturally take place. These effects should depend
on the orientation of the non-equilibrium drive with respect to the
Fermi surface, opening new possibilities for novel patterned phases.
Non-equilibrium phase transitions to patterned phases, in particular
at zero temperature where quantum effects are most relevant, present
an interesting paradigm where new universal behavior could be found.
\begin{acknowledgments}
AEA acknowledges Russian Quantum Center for hospitality.
\end{acknowledgments}

\bibliographystyle{apsrev4-1}
\bibliography{OrderNESS2}

\begin{thebibliography}{49}%
\makeatletter
\providecommand \@ifxundefined [1]{%
 \@ifx{#1\undefined}
}%
\providecommand \@ifnum [1]{%
 \ifnum #1\expandafter \@firstoftwo
 \else \expandafter \@secondoftwo
 \fi
}%
\providecommand \@ifx [1]{%
 \ifx #1\expandafter \@firstoftwo
 \else \expandafter \@secondoftwo
 \fi
}%
\providecommand \natexlab [1]{#1}%
\providecommand \enquote  [1]{``#1''}%
\providecommand \bibnamefont  [1]{#1}%
\providecommand \bibfnamefont [1]{#1}%
\providecommand \citenamefont [1]{#1}%
\providecommand \href@noop [0]{\@secondoftwo}%
\providecommand \href [0]{\begingroup \@sanitize@url \@href}%
\providecommand \@href[1]{\@@startlink{#1}\@@href}%
\providecommand \@@href[1]{\endgroup#1\@@endlink}%
\providecommand \@sanitize@url [0]{\catcode `\\12\catcode `\$12\catcode
  `\&12\catcode `\#12\catcode `\^12\catcode `\_12\catcode `\%12\relax}%
\providecommand \@@startlink[1]{}%
\providecommand \@@endlink[0]{}%
\providecommand \url  [0]{\begingroup\@sanitize@url \@url }%
\providecommand \@url [1]{\endgroup\@href {#1}{\urlprefix }}%
\providecommand \urlprefix  [0]{URL }%
\providecommand \Eprint [0]{\href }%
\providecommand \doibase [0]{http://dx.doi.org/}%
\providecommand \selectlanguage [0]{\@gobble}%
\providecommand \bibinfo  [0]{\@secondoftwo}%
\providecommand \bibfield  [0]{\@secondoftwo}%
\providecommand \translation [1]{[#1]}%
\providecommand \BibitemOpen [0]{}%
\providecommand \bibitemStop [0]{}%
\providecommand \bibitemNoStop [0]{.\EOS\space}%
\providecommand \EOS [0]{\spacefactor3000\relax}%
\providecommand \BibitemShut  [1]{\csname bibitem#1\endcsname}%
\let\auto@bib@innerbib\@empty
%</preamble>
\bibitem [{\citenamefont {Cross}\ and\ \citenamefont
  {Hohenberg}(1993)}]{Cross1993}%
  \BibitemOpen
  \bibfield  {author} {\bibinfo {author} {\bibfnamefont {M.}~\bibnamefont
  {Cross}}\ and\ \bibinfo {author} {\bibfnamefont {P.}~\bibnamefont
  {Hohenberg}},\ }\href {\doibase 10.1103/RevModPhys.65.851} {\bibfield
  {journal} {\bibinfo  {journal} {Reviews of Modern Physics}\ }\textbf
  {\bibinfo {volume} {65}},\ \bibinfo {pages} {851} (\bibinfo {year}
  {1993})}\BibitemShut {NoStop}%
\bibitem [{\citenamefont {Nicolis}\ and\ \citenamefont
  {Prigogine}(1977)}]{Nicolis1977}%
  \BibitemOpen
  \bibfield  {author} {\bibinfo {author} {\bibfnamefont {G.}~\bibnamefont
  {Nicolis}}\ and\ \bibinfo {author} {\bibfnamefont {I.}~\bibnamefont
  {Prigogine}},\ }\href@noop {} {\emph {\bibinfo {title} {{Self-organization in
  nonequilibrium systems: from dissipative structures to order through
  fluctuations}}}}\ (\bibinfo  {publisher} {Wiley},\ \bibinfo {year} {1977})\
  p.\ \bibinfo {pages} {491}\BibitemShut {NoStop}%
\bibitem [{\citenamefont {Ball}(1999)}]{Ball}%
  \BibitemOpen
  \bibfield  {author} {\bibinfo {author} {\bibfnamefont {P.}~\bibnamefont
  {Ball}},\ }\href@noop {} {\emph {\bibinfo {title} {{The Self-Made Tapestry:
  Pattern Formation in Nature}}}}\ (\bibinfo  {publisher} {Oxford University
  Press},\ \bibinfo {year} {1999})\BibitemShut {NoStop}%
\bibitem [{\citenamefont {Sch\"{o}ll}(1987)}]{Scholl1987}%
  \BibitemOpen
  \bibfield  {author} {\bibinfo {author} {\bibfnamefont {E.}~\bibnamefont
  {Sch\"{o}ll}},\ }\href {\doibase 10.1007/978-3-642-71927-1} {\emph {\bibinfo
  {title} {{Nonequilibrium Phase Transitions in Semiconductors}}}},\ \bibinfo
  {series} {Springer Series in Synergetics}, Vol.~\bibinfo {volume} {35}\
  (\bibinfo  {publisher} {Springer Berlin Heidelberg},\ \bibinfo {address}
  {Berlin, Heidelberg},\ \bibinfo {year} {1987})\BibitemShut {NoStop}%
\bibitem [{\citenamefont {Kumai}(1999)}]{Kumai1999}%
  \BibitemOpen
  \bibfield  {author} {\bibinfo {author} {\bibfnamefont {R.}~\bibnamefont
  {Kumai}},\ }\href {\doibase 10.1126/science.284.5420.1645} {\bibfield
  {journal} {\bibinfo  {journal} {Science}\ }\textbf {\bibinfo {volume}
  {284}},\ \bibinfo {pages} {1645} (\bibinfo {year} {1999})}\BibitemShut
  {NoStop}%
\bibitem [{\citenamefont {Kronj\"{a}ger}\ \emph {et~al.}(2010)\citenamefont
  {Kronj\"{a}ger}, \citenamefont {Becker}, \citenamefont {Soltan-Panahi},
  \citenamefont {Bongs},\ and\ \citenamefont {Sengstock}}]{Kronjager2010}%
  \BibitemOpen
  \bibfield  {author} {\bibinfo {author} {\bibfnamefont {J.}~\bibnamefont
  {Kronj\"{a}ger}}, \bibinfo {author} {\bibfnamefont {C.}~\bibnamefont
  {Becker}}, \bibinfo {author} {\bibfnamefont {P.}~\bibnamefont
  {Soltan-Panahi}}, \bibinfo {author} {\bibfnamefont {K.}~\bibnamefont
  {Bongs}}, \ and\ \bibinfo {author} {\bibfnamefont {K.}~\bibnamefont
  {Sengstock}},\ }\href {\doibase 10.1103/PhysRevLett.105.090402} {\bibfield
  {journal} {\bibinfo  {journal} {Physical Review Letters}\ }\textbf {\bibinfo
  {volume} {105}},\ \bibinfo {pages} {090402} (\bibinfo {year}
  {2010})}\BibitemShut {NoStop}%
\bibitem [{\citenamefont {Borgh}\ \emph {et~al.}(2010)\citenamefont {Borgh},
  \citenamefont {Keeling},\ and\ \citenamefont {Berloff}}]{Borgh2010}%
  \BibitemOpen
  \bibfield  {author} {\bibinfo {author} {\bibfnamefont {M.~O.}\ \bibnamefont
  {Borgh}}, \bibinfo {author} {\bibfnamefont {J.}~\bibnamefont {Keeling}}, \
  and\ \bibinfo {author} {\bibfnamefont {N.~G.}\ \bibnamefont {Berloff}},\
  }\href {\doibase 10.1103/PhysRevB.81.235302} {\bibfield  {journal} {\bibinfo
  {journal} {Physical Review B}\ }\textbf {\bibinfo {volume} {81}},\ \bibinfo
  {pages} {235302} (\bibinfo {year} {2010})}\BibitemShut {NoStop}%
\bibitem [{\citenamefont {Berloff}\ and\ \citenamefont
  {Keeling}(2013)}]{Berloff2013}%
  \BibitemOpen
  \bibfield  {author} {\bibinfo {author} {\bibfnamefont {N.}~\bibnamefont
  {Berloff}}\ and\ \bibinfo {author} {\bibfnamefont {J.}~\bibnamefont
  {Keeling}},\ }\href {\doibase 10.1007/978-3-642-37569-9} {\emph {\bibinfo
  {title} {{Physics of Quantum Fluids}}}},\ edited by\ \bibinfo {editor}
  {\bibfnamefont {A.}~\bibnamefont {Bramati}}\ and\ \bibinfo {editor}
  {\bibfnamefont {M.}~\bibnamefont {Modugno}},\ \bibinfo {series} {Springer
  Series in Solid-State Sciences}, Vol.\ \bibinfo {volume} {177}\ (\bibinfo
  {publisher} {Springer Berlin Heidelberg},\ \bibinfo {address} {Berlin,
  Heidelberg},\ \bibinfo {year} {2013})\BibitemShut {NoStop}%
\bibitem [{\citenamefont {Bloch}\ and\ \citenamefont
  {Zwerger}(2008)}]{Bloch2008}%
  \BibitemOpen
  \bibfield  {author} {\bibinfo {author} {\bibfnamefont {I.}~\bibnamefont
  {Bloch}}\ and\ \bibinfo {author} {\bibfnamefont {W.}~\bibnamefont
  {Zwerger}},\ }\href {\doibase 10.1103/RevModPhys.80.885} {\bibfield
  {journal} {\bibinfo  {journal} {Reviews of Modern Physics}\ }\textbf
  {\bibinfo {volume} {80}},\ \bibinfo {pages} {885} (\bibinfo {year}
  {2008})}\BibitemShut {NoStop}%
\bibitem [{\citenamefont {Strohmaier}\ \emph {et~al.}(2010)\citenamefont
  {Strohmaier}, \citenamefont {Greif}, \citenamefont {J\"{o}rdens},
  \citenamefont {Tarruell}, \citenamefont {Moritz}, \citenamefont {Esslinger},
  \citenamefont {Sensarma}, \citenamefont {Pekker}, \citenamefont {Altman},\
  and\ \citenamefont {Demler}}]{Strohmaier2010}%
  \BibitemOpen
  \bibfield  {author} {\bibinfo {author} {\bibfnamefont {N.}~\bibnamefont
  {Strohmaier}}, \bibinfo {author} {\bibfnamefont {D.}~\bibnamefont {Greif}},
  \bibinfo {author} {\bibfnamefont {R.}~\bibnamefont {J\"{o}rdens}}, \bibinfo
  {author} {\bibfnamefont {L.}~\bibnamefont {Tarruell}}, \bibinfo {author}
  {\bibfnamefont {H.}~\bibnamefont {Moritz}}, \bibinfo {author} {\bibfnamefont
  {T.}~\bibnamefont {Esslinger}}, \bibinfo {author} {\bibfnamefont
  {R.}~\bibnamefont {Sensarma}}, \bibinfo {author} {\bibfnamefont
  {D.}~\bibnamefont {Pekker}}, \bibinfo {author} {\bibfnamefont
  {E.}~\bibnamefont {Altman}}, \ and\ \bibinfo {author} {\bibfnamefont
  {E.}~\bibnamefont {Demler}},\ }\href
  {http://link.aps.org/doi/10.1103/PhysRevLett.104.080401} {\bibfield
  {journal} {\bibinfo  {journal} {Physical Review Letters}\ }\textbf {\bibinfo
  {volume} {104}},\ \bibinfo {pages} {080401} (\bibinfo {year}
  {2010})}\BibitemShut {NoStop}%
\bibitem [{\citenamefont {Cavalleri}\ \emph {et~al.}(2001)\citenamefont
  {Cavalleri}, \citenamefont {T\'{o}th}, \citenamefont {Siders}, \citenamefont
  {Squier}, \citenamefont {R\'{a}ksi}, \citenamefont {Forget},\ and\
  \citenamefont {Kieffer}}]{Cavalleri2001}%
  \BibitemOpen
  \bibfield  {author} {\bibinfo {author} {\bibfnamefont {A.}~\bibnamefont
  {Cavalleri}}, \bibinfo {author} {\bibfnamefont {C.}~\bibnamefont {T\'{o}th}},
  \bibinfo {author} {\bibfnamefont {C.}~\bibnamefont {Siders}}, \bibinfo
  {author} {\bibfnamefont {J.}~\bibnamefont {Squier}}, \bibinfo {author}
  {\bibfnamefont {F.}~\bibnamefont {R\'{a}ksi}}, \bibinfo {author}
  {\bibfnamefont {P.}~\bibnamefont {Forget}}, \ and\ \bibinfo {author}
  {\bibfnamefont {J.}~\bibnamefont {Kieffer}},\ }\href {\doibase
  10.1103/PhysRevLett.87.237401} {\bibfield  {journal} {\bibinfo  {journal}
  {Physical Review Letters}\ }\textbf {\bibinfo {volume} {87}},\ \bibinfo
  {pages} {237401} (\bibinfo {year} {2001})}\BibitemShut {NoStop}%
\bibitem [{\citenamefont {Novelli}\ \emph {et~al.}(2014)\citenamefont
  {Novelli}, \citenamefont {{De Filippis}}, \citenamefont {Cataudella},
  \citenamefont {Esposito}, \citenamefont {Vergara}, \citenamefont {Cilento},
  \citenamefont {Sindici}, \citenamefont {Amaricci}, \citenamefont {Giannetti},
  \citenamefont {Prabhakaran}, \citenamefont {Wall}, \citenamefont {Perucchi},
  \citenamefont {{Dal Conte}}, \citenamefont {Cerullo}, \citenamefont {Capone},
  \citenamefont {Mishchenko}, \citenamefont {Gr\"{u}ninger}, \citenamefont
  {Nagaosa}, \citenamefont {Parmigiani},\ and\ \citenamefont
  {Fausti}}]{Novelli2014}%
  \BibitemOpen
  \bibfield  {author} {\bibinfo {author} {\bibfnamefont {F.}~\bibnamefont
  {Novelli}}, \bibinfo {author} {\bibfnamefont {G.}~\bibnamefont {{De
  Filippis}}}, \bibinfo {author} {\bibfnamefont {V.}~\bibnamefont
  {Cataudella}}, \bibinfo {author} {\bibfnamefont {M.}~\bibnamefont
  {Esposito}}, \bibinfo {author} {\bibfnamefont {I.}~\bibnamefont {Vergara}},
  \bibinfo {author} {\bibfnamefont {F.}~\bibnamefont {Cilento}}, \bibinfo
  {author} {\bibfnamefont {E.}~\bibnamefont {Sindici}}, \bibinfo {author}
  {\bibfnamefont {A.}~\bibnamefont {Amaricci}}, \bibinfo {author}
  {\bibfnamefont {C.}~\bibnamefont {Giannetti}}, \bibinfo {author}
  {\bibfnamefont {D.}~\bibnamefont {Prabhakaran}}, \bibinfo {author}
  {\bibfnamefont {S.}~\bibnamefont {Wall}}, \bibinfo {author} {\bibfnamefont
  {A.}~\bibnamefont {Perucchi}}, \bibinfo {author} {\bibfnamefont
  {S.}~\bibnamefont {{Dal Conte}}}, \bibinfo {author} {\bibfnamefont
  {G.}~\bibnamefont {Cerullo}}, \bibinfo {author} {\bibfnamefont
  {M.}~\bibnamefont {Capone}}, \bibinfo {author} {\bibfnamefont
  {A.}~\bibnamefont {Mishchenko}}, \bibinfo {author} {\bibfnamefont
  {M.}~\bibnamefont {Gr\"{u}ninger}}, \bibinfo {author} {\bibfnamefont
  {N.}~\bibnamefont {Nagaosa}}, \bibinfo {author} {\bibfnamefont
  {F.}~\bibnamefont {Parmigiani}}, \ and\ \bibinfo {author} {\bibfnamefont
  {D.}~\bibnamefont {Fausti}},\ }\href {\doibase 10.1038/ncomms6112} {\bibfield
   {journal} {\bibinfo  {journal} {Nature communications}\ }\textbf {\bibinfo
  {volume} {5}},\ \bibinfo {pages} {5112} (\bibinfo {year} {2014})}\BibitemShut
  {NoStop}%
\bibitem [{\citenamefont {Rigol}\ \emph {et~al.}(2008)\citenamefont {Rigol},
  \citenamefont {Dunjko},\ and\ \citenamefont {Olshanii}}]{Rigol2008b}%
  \BibitemOpen
  \bibfield  {author} {\bibinfo {author} {\bibfnamefont {M.}~\bibnamefont
  {Rigol}}, \bibinfo {author} {\bibfnamefont {V.}~\bibnamefont {Dunjko}}, \
  and\ \bibinfo {author} {\bibfnamefont {M.}~\bibnamefont {Olshanii}},\ }\href
  {\doibase 10.1038/nature06838} {\bibfield  {journal} {\bibinfo  {journal}
  {Nature}\ }\textbf {\bibinfo {volume} {452}},\ \bibinfo {pages} {854}
  (\bibinfo {year} {2008})}\BibitemShut {NoStop}%
\bibitem [{\citenamefont {Srednicki}(1994)}]{Srednicki1994}%
  \BibitemOpen
  \bibfield  {author} {\bibinfo {author} {\bibfnamefont {M.}~\bibnamefont
  {Srednicki}},\ }\href {\doibase 10.1103/PhysRevE.50.888} {\bibfield
  {journal} {\bibinfo  {journal} {Physical Review E}\ }\textbf {\bibinfo
  {volume} {50}},\ \bibinfo {pages} {888} (\bibinfo {year} {1994})}\BibitemShut
  {NoStop}%
\bibitem [{\citenamefont {Deutsch}(1991)}]{Deutsch1991}%
  \BibitemOpen
  \bibfield  {author} {\bibinfo {author} {\bibfnamefont {J.}~\bibnamefont
  {Deutsch}},\ }\href {\doibase 10.1103/PhysRevA.43.2046} {\bibfield  {journal}
  {\bibinfo  {journal} {Physical Review A}\ }\textbf {\bibinfo {volume} {43}},\
  \bibinfo {pages} {2046} (\bibinfo {year} {1991})}\BibitemShut {NoStop}%
\bibitem [{\citenamefont {Diehl}\ \emph {et~al.}(2008)\citenamefont {Diehl},
  \citenamefont {Micheli}, \citenamefont {Kantian}, \citenamefont {Kraus},
  \citenamefont {B\"{u}chler},\ and\ \citenamefont {Zoller}}]{Diehl2008}%
  \BibitemOpen
  \bibfield  {author} {\bibinfo {author} {\bibfnamefont {S.}~\bibnamefont
  {Diehl}}, \bibinfo {author} {\bibfnamefont {a.}~\bibnamefont {Micheli}},
  \bibinfo {author} {\bibfnamefont {a.}~\bibnamefont {Kantian}}, \bibinfo
  {author} {\bibfnamefont {B.}~\bibnamefont {Kraus}}, \bibinfo {author}
  {\bibfnamefont {H.~P.}\ \bibnamefont {B\"{u}chler}}, \ and\ \bibinfo {author}
  {\bibfnamefont {P.}~\bibnamefont {Zoller}},\ }\href {\doibase
  10.1038/nphys1073} {\bibfield  {journal} {\bibinfo  {journal} {Nature
  Physics}\ }\textbf {\bibinfo {volume} {4}},\ \bibinfo {pages} {878} (\bibinfo
  {year} {2008})}\BibitemShut {NoStop}%
\bibitem [{\citenamefont {Diehl}\ \emph {et~al.}(2010)\citenamefont {Diehl},
  \citenamefont {Tomadin}, \citenamefont {Micheli}, \citenamefont {Fazio},\
  and\ \citenamefont {Zoller}}]{Diehl2010}%
  \BibitemOpen
  \bibfield  {author} {\bibinfo {author} {\bibfnamefont {S.}~\bibnamefont
  {Diehl}}, \bibinfo {author} {\bibfnamefont {A.}~\bibnamefont {Tomadin}},
  \bibinfo {author} {\bibfnamefont {A.}~\bibnamefont {Micheli}}, \bibinfo
  {author} {\bibfnamefont {R.}~\bibnamefont {Fazio}}, \ and\ \bibinfo {author}
  {\bibfnamefont {P.}~\bibnamefont {Zoller}},\ }\href {\doibase
  10.1103/PhysRevLett.105.015702} {\bibfield  {journal} {\bibinfo  {journal}
  {Physical Review Letters}\ }\textbf {\bibinfo {volume} {105}},\ \bibinfo
  {pages} {015702} (\bibinfo {year} {2010})}\BibitemShut {NoStop}%
\bibitem [{\citenamefont {Sieberer}\ \emph {et~al.}(2013)\citenamefont
  {Sieberer}, \citenamefont {Huber}, \citenamefont {Altman},\ and\
  \citenamefont {Diehl}}]{Sieberer2013}%
  \BibitemOpen
  \bibfield  {author} {\bibinfo {author} {\bibfnamefont {L.~M.}\ \bibnamefont
  {Sieberer}}, \bibinfo {author} {\bibfnamefont {S.~D.}\ \bibnamefont {Huber}},
  \bibinfo {author} {\bibfnamefont {E.}~\bibnamefont {Altman}}, \ and\ \bibinfo
  {author} {\bibfnamefont {S.}~\bibnamefont {Diehl}},\ }\href {\doibase
  10.1103/PhysRevLett.110.195301} {\bibfield  {journal} {\bibinfo  {journal}
  {Physical Review Letters}\ }\textbf {\bibinfo {volume} {110}},\ \bibinfo
  {pages} {195301} (\bibinfo {year} {2013})}\BibitemShut {NoStop}%
\bibitem [{\citenamefont {Mitra}\ \emph {et~al.}(2006)\citenamefont {Mitra},
  \citenamefont {Takei}, \citenamefont {Kim},\ and\ \citenamefont
  {Millis}}]{Mitra2006}%
  \BibitemOpen
  \bibfield  {author} {\bibinfo {author} {\bibfnamefont {A.}~\bibnamefont
  {Mitra}}, \bibinfo {author} {\bibfnamefont {S.}~\bibnamefont {Takei}},
  \bibinfo {author} {\bibfnamefont {Y.}~\bibnamefont {Kim}}, \ and\ \bibinfo
  {author} {\bibfnamefont {A.}~\bibnamefont {Millis}},\ }\href {\doibase
  10.1103/PhysRevLett.97.236808} {\bibfield  {journal} {\bibinfo  {journal}
  {Physical Review Letters}\ }\textbf {\bibinfo {volume} {97}},\ \bibinfo
  {pages} {236808} (\bibinfo {year} {2006})}\BibitemShut {NoStop}%
\bibitem [{\citenamefont {Mitra}\ and\ \citenamefont
  {Millis}(2008)}]{Mitra2008a}%
  \BibitemOpen
  \bibfield  {author} {\bibinfo {author} {\bibfnamefont {A.}~\bibnamefont
  {Mitra}}\ and\ \bibinfo {author} {\bibfnamefont {A.}~\bibnamefont {Millis}},\
  }\href {\doibase 10.1103/PhysRevB.77.220404} {\bibfield  {journal} {\bibinfo
  {journal} {Physical Review B}\ }\textbf {\bibinfo {volume} {77}},\ \bibinfo
  {pages} {220404} (\bibinfo {year} {2008})}\BibitemShut {NoStop}%
\bibitem [{\citenamefont {Takei}\ \emph {et~al.}(2010)\citenamefont {Takei},
  \citenamefont {Witczak-Krempa},\ and\ \citenamefont {Kim}}]{Takei2010}%
  \BibitemOpen
  \bibfield  {author} {\bibinfo {author} {\bibfnamefont {S.}~\bibnamefont
  {Takei}}, \bibinfo {author} {\bibfnamefont {W.}~\bibnamefont
  {Witczak-Krempa}}, \ and\ \bibinfo {author} {\bibfnamefont {Y.~B.}\
  \bibnamefont {Kim}},\ }\href {\doibase 10.1103/PhysRevB.81.125430} {\bibfield
   {journal} {\bibinfo  {journal} {Physical Review B}\ }\textbf {\bibinfo
  {volume} {81}},\ \bibinfo {pages} {125430} (\bibinfo {year}
  {2010})}\BibitemShut {NoStop}%
\bibitem [{\citenamefont {Chung}\ \emph {et~al.}(2009)\citenamefont {Chung},
  \citenamefont {{Le Hur}}, \citenamefont {Vojta},\ and\ \citenamefont
  {W\"{o}lfle}}]{Chung2009}%
  \BibitemOpen
  \bibfield  {author} {\bibinfo {author} {\bibfnamefont {C.-H.}\ \bibnamefont
  {Chung}}, \bibinfo {author} {\bibfnamefont {K.}~\bibnamefont {{Le Hur}}},
  \bibinfo {author} {\bibfnamefont {M.}~\bibnamefont {Vojta}}, \ and\ \bibinfo
  {author} {\bibfnamefont {P.}~\bibnamefont {W\"{o}lfle}},\ }\href {\doibase
  10.1103/PhysRevLett.102.216803} {\bibfield  {journal} {\bibinfo  {journal}
  {Physical Review Letters}\ }\textbf {\bibinfo {volume} {102}},\ \bibinfo
  {pages} {216803} (\bibinfo {year} {2009})}\BibitemShut {NoStop}%
\bibitem [{\citenamefont {Kirchner}\ and\ \citenamefont
  {Si}(2009)}]{Kirchner2009a}%
  \BibitemOpen
  \bibfield  {author} {\bibinfo {author} {\bibfnamefont {S.}~\bibnamefont
  {Kirchner}}\ and\ \bibinfo {author} {\bibfnamefont {Q.}~\bibnamefont {Si}},\
  }\href {\doibase 10.1103/PhysRevLett.103.206401} {\bibfield  {journal}
  {\bibinfo  {journal} {Physical Review Letters}\ }\textbf {\bibinfo {volume}
  {103}},\ \bibinfo {pages} {206401} (\bibinfo {year} {2009})}\BibitemShut
  {NoStop}%
\bibitem [{\citenamefont {Ribeiro}\ \emph {et~al.}(2013)\citenamefont
  {Ribeiro}, \citenamefont {Si},\ and\ \citenamefont
  {Kirchner}}]{Ribeiro2013b}%
  \BibitemOpen
  \bibfield  {author} {\bibinfo {author} {\bibfnamefont {P.}~\bibnamefont
  {Ribeiro}}, \bibinfo {author} {\bibfnamefont {Q.}~\bibnamefont {Si}}, \ and\
  \bibinfo {author} {\bibfnamefont {S.}~\bibnamefont {Kirchner}},\ }\href
  {\doibase 10.1209/0295-5075/102/50001} {\bibfield  {journal} {\bibinfo
  {journal} {EPL (Europhysics Letters)}\ }\textbf {\bibinfo {volume} {102}},\
  \bibinfo {pages} {50001} (\bibinfo {year} {2013})}\BibitemShut {NoStop}%
\bibitem [{\citenamefont {Werner}\ \emph {et~al.}(2009)\citenamefont {Werner},
  \citenamefont {Oka},\ and\ \citenamefont {Millis}}]{Werner2009}%
  \BibitemOpen
  \bibfield  {author} {\bibinfo {author} {\bibfnamefont {P.}~\bibnamefont
  {Werner}}, \bibinfo {author} {\bibfnamefont {T.}~\bibnamefont {Oka}}, \ and\
  \bibinfo {author} {\bibfnamefont {A.}~\bibnamefont {Millis}},\ }\href
  {\doibase 10.1103/PhysRevB.79.035320} {\bibfield  {journal} {\bibinfo
  {journal} {Physical Review B}\ }\textbf {\bibinfo {volume} {79}},\ \bibinfo
  {pages} {035320} (\bibinfo {year} {2009})}\BibitemShut {NoStop}%
\bibitem [{\citenamefont {Schir\'{o}}\ and\ \citenamefont
  {Fabrizio}(2009)}]{Schiro2009}%
  \BibitemOpen
  \bibfield  {author} {\bibinfo {author} {\bibfnamefont {M.}~\bibnamefont
  {Schir\'{o}}}\ and\ \bibinfo {author} {\bibfnamefont {M.}~\bibnamefont
  {Fabrizio}},\ }\href {\doibase 10.1103/PhysRevB.79.153302} {\bibfield
  {journal} {\bibinfo  {journal} {Physical Review B}\ }\textbf {\bibinfo
  {volume} {79}},\ \bibinfo {pages} {153302} (\bibinfo {year}
  {2009})}\BibitemShut {NoStop}%
\bibitem [{\citenamefont {Gull}\ \emph {et~al.}(2010)\citenamefont {Gull},
  \citenamefont {Reichman},\ and\ \citenamefont {Millis}}]{Gull2010}%
  \BibitemOpen
  \bibfield  {author} {\bibinfo {author} {\bibfnamefont {E.}~\bibnamefont
  {Gull}}, \bibinfo {author} {\bibfnamefont {D.~R.}\ \bibnamefont {Reichman}},
  \ and\ \bibinfo {author} {\bibfnamefont {A.~J.}\ \bibnamefont {Millis}},\
  }\href {\doibase 10.1103/PhysRevB.82.075109} {\bibfield  {journal} {\bibinfo
  {journal} {Physical Review B}\ }\textbf {\bibinfo {volume} {82}},\ \bibinfo
  {pages} {075109} (\bibinfo {year} {2010})}\BibitemShut {NoStop}%
\bibitem [{\citenamefont {Gull}\ \emph {et~al.}(2011)\citenamefont {Gull},
  \citenamefont {Reichman},\ and\ \citenamefont {Millis}}]{Gull2011}%
  \BibitemOpen
  \bibfield  {author} {\bibinfo {author} {\bibfnamefont {E.}~\bibnamefont
  {Gull}}, \bibinfo {author} {\bibfnamefont {D.~R.}\ \bibnamefont {Reichman}},
  \ and\ \bibinfo {author} {\bibfnamefont {A.~J.}\ \bibnamefont {Millis}},\
  }\href {\doibase 10.1103/PhysRevB.84.085134} {\bibfield  {journal} {\bibinfo
  {journal} {Physical Review B}\ }\textbf {\bibinfo {volume} {84}},\ \bibinfo
  {pages} {085134} (\bibinfo {year} {2011})}\BibitemShut {NoStop}%
\bibitem [{\citenamefont {Cohen}\ \emph {et~al.}(2013)\citenamefont {Cohen},
  \citenamefont {Gull}, \citenamefont {Reichman}, \citenamefont {Millis},\ and\
  \citenamefont {Rabani}}]{Cohen2013}%
  \BibitemOpen
  \bibfield  {author} {\bibinfo {author} {\bibfnamefont {G.}~\bibnamefont
  {Cohen}}, \bibinfo {author} {\bibfnamefont {E.}~\bibnamefont {Gull}},
  \bibinfo {author} {\bibfnamefont {D.~R.}\ \bibnamefont {Reichman}}, \bibinfo
  {author} {\bibfnamefont {A.~J.}\ \bibnamefont {Millis}}, \ and\ \bibinfo
  {author} {\bibfnamefont {E.}~\bibnamefont {Rabani}},\ }\href {\doibase
  10.1103/PhysRevB.87.195108} {\bibfield  {journal} {\bibinfo  {journal}
  {Physical Review B}\ }\textbf {\bibinfo {volume} {87}},\ \bibinfo {pages}
  {195108} (\bibinfo {year} {2013})}\BibitemShut {NoStop}%
\bibitem [{\citenamefont {Aoki}\ \emph {et~al.}(2014)\citenamefont {Aoki},
  \citenamefont {Tsuji}, \citenamefont {Eckstein}, \citenamefont {Kollar},
  \citenamefont {Oka},\ and\ \citenamefont {Werner}}]{Aoki2014}%
  \BibitemOpen
  \bibfield  {author} {\bibinfo {author} {\bibfnamefont {H.}~\bibnamefont
  {Aoki}}, \bibinfo {author} {\bibfnamefont {N.}~\bibnamefont {Tsuji}},
  \bibinfo {author} {\bibfnamefont {M.}~\bibnamefont {Eckstein}}, \bibinfo
  {author} {\bibfnamefont {M.}~\bibnamefont {Kollar}}, \bibinfo {author}
  {\bibfnamefont {T.}~\bibnamefont {Oka}}, \ and\ \bibinfo {author}
  {\bibfnamefont {P.}~\bibnamefont {Werner}},\ }\href {\doibase
  10.1103/RevModPhys.86.779} {\bibfield  {journal} {\bibinfo  {journal}
  {Reviews of Modern Physics}\ }\textbf {\bibinfo {volume} {86}},\ \bibinfo
  {pages} {779} (\bibinfo {year} {2014})}\BibitemShut {NoStop}%
\bibitem [{\citenamefont {Schir\'{o}}\ and\ \citenamefont
  {Fabrizio}(2010)}]{Schiro2010}%
  \BibitemOpen
  \bibfield  {author} {\bibinfo {author} {\bibfnamefont {M.}~\bibnamefont
  {Schir\'{o}}}\ and\ \bibinfo {author} {\bibfnamefont {M.}~\bibnamefont
  {Fabrizio}},\ }\href {\doibase 10.1103/PhysRevLett.105.076401} {\bibfield
  {journal} {\bibinfo  {journal} {Physical Review Letters}\ }\textbf {\bibinfo
  {volume} {105}},\ \bibinfo {pages} {076401} (\bibinfo {year}
  {2010})}\BibitemShut {NoStop}%
\bibitem [{\citenamefont {Schir\'{o}}\ and\ \citenamefont
  {Fabrizio}(2011)}]{Schiro2011}%
  \BibitemOpen
  \bibfield  {author} {\bibinfo {author} {\bibfnamefont {M.}~\bibnamefont
  {Schir\'{o}}}\ and\ \bibinfo {author} {\bibfnamefont {M.}~\bibnamefont
  {Fabrizio}},\ }\href {\doibase 10.1103/PhysRevB.83.165105} {\bibfield
  {journal} {\bibinfo  {journal} {Physical Review B}\ }\textbf {\bibinfo
  {volume} {83}},\ \bibinfo {pages} {165105} (\bibinfo {year}
  {2011})}\BibitemShut {NoStop}%
\bibitem [{\citenamefont {Eckstein}\ \emph
  {et~al.}(2010{\natexlab{a}})\citenamefont {Eckstein}, \citenamefont {Hackl},
  \citenamefont {Kehrein}, \citenamefont {Kollar}, \citenamefont {Moeckel},
  \citenamefont {Werner},\ and\ \citenamefont {Wolf}}]{Eckstein2010}%
  \BibitemOpen
  \bibfield  {author} {\bibinfo {author} {\bibfnamefont {M.}~\bibnamefont
  {Eckstein}}, \bibinfo {author} {\bibfnamefont {A.}~\bibnamefont {Hackl}},
  \bibinfo {author} {\bibfnamefont {S.}~\bibnamefont {Kehrein}}, \bibinfo
  {author} {\bibfnamefont {M.}~\bibnamefont {Kollar}}, \bibinfo {author}
  {\bibfnamefont {M.}~\bibnamefont {Moeckel}}, \bibinfo {author} {\bibfnamefont
  {P.}~\bibnamefont {Werner}}, \ and\ \bibinfo {author} {\bibfnamefont
  {F.}~\bibnamefont {Wolf}},\ }\href {\doibase 10.1140/epjst/e2010-01219-x}
  {\bibfield  {journal} {\bibinfo  {journal} {The European Physical Journal
  Special Topics}\ }\textbf {\bibinfo {volume} {180}},\ \bibinfo {pages} {217}
  (\bibinfo {year} {2010}{\natexlab{a}})}\BibitemShut {NoStop}%
\bibitem [{\citenamefont {Aron}\ \emph {et~al.}(2012)\citenamefont {Aron},
  \citenamefont {Kotliar},\ and\ \citenamefont {Weber}}]{Aron2012a}%
  \BibitemOpen
  \bibfield  {author} {\bibinfo {author} {\bibfnamefont {C.}~\bibnamefont
  {Aron}}, \bibinfo {author} {\bibfnamefont {G.}~\bibnamefont {Kotliar}}, \
  and\ \bibinfo {author} {\bibfnamefont {C.}~\bibnamefont {Weber}},\ }\href
  {\doibase 10.1103/PhysRevLett.108.086401} {\bibfield  {journal} {\bibinfo
  {journal} {Physical Review Letters}\ }\textbf {\bibinfo {volume} {108}},\
  \bibinfo {pages} {086401} (\bibinfo {year} {2012})}\BibitemShut {NoStop}%
\bibitem [{\citenamefont {Arrigoni}\ \emph {et~al.}(2013)\citenamefont
  {Arrigoni}, \citenamefont {Knap},\ and\ \citenamefont {von~der
  Linden}}]{Arrigoni2013}%
  \BibitemOpen
  \bibfield  {author} {\bibinfo {author} {\bibfnamefont {E.}~\bibnamefont
  {Arrigoni}}, \bibinfo {author} {\bibfnamefont {M.}~\bibnamefont {Knap}}, \
  and\ \bibinfo {author} {\bibfnamefont {W.}~\bibnamefont {von~der Linden}},\
  }\href {\doibase 10.1103/PhysRevLett.110.086403} {\bibfield  {journal}
  {\bibinfo  {journal} {Physical Review Letters}\ }\textbf {\bibinfo {volume}
  {110}},\ \bibinfo {pages} {086403} (\bibinfo {year} {2013})}\BibitemShut
  {NoStop}%
\bibitem [{\citenamefont {Moeckel}\ and\ \citenamefont
  {Kehrein}(2008)}]{Moeckel2008}%
  \BibitemOpen
  \bibfield  {author} {\bibinfo {author} {\bibfnamefont {M.}~\bibnamefont
  {Moeckel}}\ and\ \bibinfo {author} {\bibfnamefont {S.}~\bibnamefont
  {Kehrein}},\ }\href {\doibase 10.1103/PhysRevLett.100.175702} {\bibfield
  {journal} {\bibinfo  {journal} {Physical Review Letters}\ }\textbf {\bibinfo
  {volume} {100}},\ \bibinfo {pages} {175702} (\bibinfo {year}
  {2008})}\BibitemShut {NoStop}%
\bibitem [{\citenamefont {Eckstein}\ \emph {et~al.}(2009)\citenamefont
  {Eckstein}, \citenamefont {Kollar},\ and\ \citenamefont
  {Werner}}]{Eckstein2009}%
  \BibitemOpen
  \bibfield  {author} {\bibinfo {author} {\bibfnamefont {M.}~\bibnamefont
  {Eckstein}}, \bibinfo {author} {\bibfnamefont {M.}~\bibnamefont {Kollar}}, \
  and\ \bibinfo {author} {\bibfnamefont {P.}~\bibnamefont {Werner}},\ }\href
  {\doibase 10.1103/PhysRevLett.103.056403} {\bibfield  {journal} {\bibinfo
  {journal} {Physical Review Letters}\ }\textbf {\bibinfo {volume} {103}},\
  \bibinfo {pages} {056403} (\bibinfo {year} {2009})}\BibitemShut {NoStop}%
\bibitem [{\citenamefont {Enss}\ and\ \citenamefont {Sirker}(2012)}]{Enss2012}%
  \BibitemOpen
  \bibfield  {author} {\bibinfo {author} {\bibfnamefont {T.}~\bibnamefont
  {Enss}}\ and\ \bibinfo {author} {\bibfnamefont {J.}~\bibnamefont {Sirker}},\
  }\href {\doibase 10.1088/1367-2630/14/2/023008} {\bibfield  {journal}
  {\bibinfo  {journal} {New Journal of Physics}\ }\textbf {\bibinfo {volume}
  {14}},\ \bibinfo {pages} {023008} (\bibinfo {year} {2012})}\BibitemShut
  {NoStop}%
\bibitem [{\citenamefont {Karrasch}\ \emph {et~al.}(2014)\citenamefont
  {Karrasch}, \citenamefont {Kennes},\ and\ \citenamefont
  {Moore}}]{Karrasch2014}%
  \BibitemOpen
  \bibfield  {author} {\bibinfo {author} {\bibfnamefont {C.}~\bibnamefont
  {Karrasch}}, \bibinfo {author} {\bibfnamefont {D.~M.}\ \bibnamefont
  {Kennes}}, \ and\ \bibinfo {author} {\bibfnamefont {J.~E.}\ \bibnamefont
  {Moore}},\ }\href {\doibase 10.1103/PhysRevB.90.155104} {\bibfield  {journal}
  {\bibinfo  {journal} {Physical Review B}\ }\textbf {\bibinfo {volume} {90}},\
  \bibinfo {pages} {155104} (\bibinfo {year} {2014})}\BibitemShut {NoStop}%
\bibitem [{\citenamefont {Prosen}\ and\ \citenamefont
  {\v{Z}nidari\v{c}}(2012)}]{Prosen2012}%
  \BibitemOpen
  \bibfield  {author} {\bibinfo {author} {\bibfnamefont {T.}~\bibnamefont
  {Prosen}}\ and\ \bibinfo {author} {\bibfnamefont {M.}~\bibnamefont
  {\v{Z}nidari\v{c}}},\ }\href {\doibase 10.1103/PhysRevB.86.125118} {\bibfield
   {journal} {\bibinfo  {journal} {Physical Review B}\ }\textbf {\bibinfo
  {volume} {86}},\ \bibinfo {pages} {125118} (\bibinfo {year}
  {2012})}\BibitemShut {NoStop}%
\bibitem [{\citenamefont {Prosen}(2014)}]{Prosen2014}%
  \BibitemOpen
  \bibfield  {author} {\bibinfo {author} {\bibfnamefont {T.}~\bibnamefont
  {Prosen}},\ }\href {\doibase 10.1103/PhysRevLett.112.030603} {\bibfield
  {journal} {\bibinfo  {journal} {Physical Review Letters}\ }\textbf {\bibinfo
  {volume} {112}},\ \bibinfo {pages} {030603} (\bibinfo {year}
  {2014})}\BibitemShut {NoStop}%
\bibitem [{\citenamefont {Oka}\ \emph {et~al.}(2003)\citenamefont {Oka},
  \citenamefont {Arita},\ and\ \citenamefont {Aoki}}]{Oka2003}%
  \BibitemOpen
  \bibfield  {author} {\bibinfo {author} {\bibfnamefont {T.}~\bibnamefont
  {Oka}}, \bibinfo {author} {\bibfnamefont {R.}~\bibnamefont {Arita}}, \ and\
  \bibinfo {author} {\bibfnamefont {H.}~\bibnamefont {Aoki}},\ }\href {\doibase
  10.1103/PhysRevLett.91.066406} {\bibfield  {journal} {\bibinfo  {journal}
  {Physical Review Letters}\ }\textbf {\bibinfo {volume} {91}},\ \bibinfo
  {pages} {066406} (\bibinfo {year} {2003})}\BibitemShut {NoStop}%
\bibitem [{\citenamefont {Oka}\ and\ \citenamefont {Aoki}(2005)}]{Oka2005}%
  \BibitemOpen
  \bibfield  {author} {\bibinfo {author} {\bibfnamefont {T.}~\bibnamefont
  {Oka}}\ and\ \bibinfo {author} {\bibfnamefont {H.}~\bibnamefont {Aoki}},\
  }\href {\doibase 10.1103/PhysRevLett.95.137601} {\bibfield  {journal}
  {\bibinfo  {journal} {Physical Review Letters}\ }\textbf {\bibinfo {volume}
  {95}},\ \bibinfo {pages} {137601} (\bibinfo {year} {2005})}\BibitemShut
  {NoStop}%
\bibitem [{\citenamefont {Eckstein}\ \emph
  {et~al.}(2010{\natexlab{b}})\citenamefont {Eckstein}, \citenamefont {Oka},\
  and\ \citenamefont {Werner}}]{Eckstein2010b}%
  \BibitemOpen
  \bibfield  {author} {\bibinfo {author} {\bibfnamefont {M.}~\bibnamefont
  {Eckstein}}, \bibinfo {author} {\bibfnamefont {T.}~\bibnamefont {Oka}}, \
  and\ \bibinfo {author} {\bibfnamefont {P.}~\bibnamefont {Werner}},\ }\href
  {\doibase 10.1103/PhysRevLett.105.146404} {\bibfield  {journal} {\bibinfo
  {journal} {Physical Review Letters}\ }\textbf {\bibinfo {volume} {105}},\
  \bibinfo {pages} {146404} (\bibinfo {year} {2010}{\natexlab{b}})}\BibitemShut
  {NoStop}%
\bibitem [{\citenamefont {Eckstein}\ and\ \citenamefont
  {Werner}(2011)}]{Eckstein2011a}%
  \BibitemOpen
  \bibfield  {author} {\bibinfo {author} {\bibfnamefont {M.}~\bibnamefont
  {Eckstein}}\ and\ \bibinfo {author} {\bibfnamefont {P.}~\bibnamefont
  {Werner}},\ }\href {\doibase 10.1103/PhysRevLett.107.186406} {\bibfield
  {journal} {\bibinfo  {journal} {Physical Review Letters}\ }\textbf {\bibinfo
  {volume} {107}},\ \bibinfo {pages} {186406} (\bibinfo {year}
  {2011})}\BibitemShut {NoStop}%
\bibitem [{\citenamefont {Sugimoto}\ \emph {et~al.}(2008)\citenamefont
  {Sugimoto}, \citenamefont {Onoda},\ and\ \citenamefont
  {Nagaosa}}]{Sugimoto2008}%
  \BibitemOpen
  \bibfield  {author} {\bibinfo {author} {\bibfnamefont {N.}~\bibnamefont
  {Sugimoto}}, \bibinfo {author} {\bibfnamefont {S.}~\bibnamefont {Onoda}}, \
  and\ \bibinfo {author} {\bibfnamefont {N.}~\bibnamefont {Nagaosa}},\ }\href
  {\doibase 10.1103/PhysRevB.78.155104} {\bibfield  {journal} {\bibinfo
  {journal} {Physical Review B}\ }\textbf {\bibinfo {volume} {78}},\ \bibinfo
  {pages} {155104} (\bibinfo {year} {2008})}\BibitemShut {NoStop}%
\bibitem [{\citenamefont {Heidrich-Meisner}\ \emph {et~al.}(2010)\citenamefont
  {Heidrich-Meisner}, \citenamefont {Gonz\'{a}lez}, \citenamefont
  {Al-Hassanieh}, \citenamefont {Feiguin}, \citenamefont {Rozenberg},\ and\
  \citenamefont {Dagotto}}]{Heidrich-Meisner2010}%
  \BibitemOpen
  \bibfield  {author} {\bibinfo {author} {\bibfnamefont {F.}~\bibnamefont
  {Heidrich-Meisner}}, \bibinfo {author} {\bibfnamefont {I.}~\bibnamefont
  {Gonz\'{a}lez}}, \bibinfo {author} {\bibfnamefont {K.~a.}\ \bibnamefont
  {Al-Hassanieh}}, \bibinfo {author} {\bibfnamefont {a.~E.}\ \bibnamefont
  {Feiguin}}, \bibinfo {author} {\bibfnamefont {M.~J.}\ \bibnamefont
  {Rozenberg}}, \ and\ \bibinfo {author} {\bibfnamefont {E.}~\bibnamefont
  {Dagotto}},\ }\href {\doibase 10.1103/PhysRevB.82.205110} {\bibfield
  {journal} {\bibinfo  {journal} {Physical Review B}\ }\textbf {\bibinfo
  {volume} {82}},\ \bibinfo {pages} {205110} (\bibinfo {year}
  {2010})}\BibitemShut {NoStop}%
\bibitem [{\citenamefont {Tanaka}\ and\ \citenamefont
  {Yonemitsu}(2011)}]{Tanaka2011}%
  \BibitemOpen
  \bibfield  {author} {\bibinfo {author} {\bibfnamefont {Y.}~\bibnamefont
  {Tanaka}}\ and\ \bibinfo {author} {\bibfnamefont {K.}~\bibnamefont
  {Yonemitsu}},\ }\href {\doibase 10.1103/PhysRevB.83.085113} {\bibfield
  {journal} {\bibinfo  {journal} {Physical Review B}\ }\textbf {\bibinfo
  {volume} {83}},\ \bibinfo {pages} {085113} (\bibinfo {year}
  {2011})}\BibitemShut {NoStop}%
\bibitem [{\citenamefont {Ribeiro}\ and\ \citenamefont
  {Vieira}()}]{RibeiroVieira2014}%
  \BibitemOpen
  \bibfield  {author} {\bibinfo {author} {\bibfnamefont {P.}~\bibnamefont
  {Ribeiro}}\ and\ \bibinfo {author} {\bibfnamefont {V.~R.}\ \bibnamefont
  {Vieira}},\ }\href@noop {} {\bibinfo  {journal} {unpublished}\ }\BibitemShut
  {NoStop}%
\end{thebibliography}%

\appendix
\cleardoublepage

\clearpage{}

\begin{widetext}

\begin{center}
\Large \bf {Supplemental Material: Mott insulator breakdown through pattern formation}\bigskip
\par\end{center}

\begin{center}
Pedro Ribeiro$^1$, Andrey E. Antipov$^2$, Alexey N. Rubtsov$^1$\medskip
\par\end{center}

\begin{center}
\it \small
$^1$Russian Quantum Center, Novaya street 100 A, Skolkovo, Moscow area, 143025 Russia\\
$^2$Department of Physics University of Michigan, Randall Laboratory, 450 Church Street, Ann Arbor, MI 48109-1040
\par\end{center}

In this supplemental material we provide some of the details of the
analytical analysis performed in the main text of the manuscript.
After deriving the Keldysh action we obtain the saddle-point equations
used in the mean-field analysis. We provide the explicit expression
for the magnetic spin susceptibility.

\section{Keldysh Action}

\subsection{Generating Functional}

The generating function in the Keldysh contour $\gamma$ is defined
as 
\begin{eqnarray}
Z & = & \int DC\, e^{i\left[C^{\dagger}g^{-1}C\right]-i\int_{\gamma}dz\,\frac{U}{2}\sum_{\bs r}\left[n_{\bs r}\left(z\right)-1\right]^{2}}
\end{eqnarray}
where $C=\left(\begin{array}{ccc}
c & d_{L} & d_{R}\end{array}\right)^{T}$ and 
\begin{eqnarray}
g^{-1} & = & \left(\begin{array}{ccc}
g_{\Sigma}^{-1} & -V_{L} & -V_{R}\\
-V_{L}^{\dagger} & g_{L}^{-1} & 0\\
-V_{R}^{\dagger} & 0 & g_{R}^{-1}
\end{array}\right)
\end{eqnarray}
is the inverse of the bare Green's function with 
\begin{eqnarray*}
g_{\text{S};\bs r,\bs r'}^{-1}\left(z,z'\right) & = & \delta\left(z-z'\right)\left(\delta_{\bs r,\bs r'}i\pd_{z}+\tilde{\bs t}_{\bs r,\bs r'}\right)\\
g_{l;\alpha,\alpha'}^{-1} & = & \delta_{\alpha,\alpha'}\delta\left(z-z'\right)\left(i\pd_{z}-\epsilon_{l,\alpha}\right)\\
V_{l;\bs r,\alpha} & = & v_{l}\delta_{\bs r,\bs r_{l}}
\end{eqnarray*}
Using the identity $\frac{U}{2}\sum_{\bs r}\left(n_{\bs r}-1\right)^{2}=-\frac{3}{4}U\left(\bs S_{\bs r}.\bs S_{\bs r}-1\right)$,
with $\bs S_{\bs r}=\frac{1}{2}c_{\bs r,s}^{\dagger}\bs{\sigma}_{ss'}c_{\bs r,s'}$,
and inserting a 3-component Hubbard-Stratonovich $\bs{\phi}$ to decouple
the interaction, one obtains, after integrating out the electronic
degrees of freedom $Z=\int D\phi\ e^{iS\left[\phi\right]}$, where
\begin{eqnarray}
S\left[\phi\right] & = & \frac{1}{2}\left(-\frac{2}{3U}\right)\sum_{\bs r}\int_{\gamma}dz\,\bs{\phi}_{\bs r}\left(z\right).\bs{\phi}_{\bs r}\left(z\right)-i\,\tr\ln\left[-iG^{-1}\right]
\end{eqnarray}
with 
\begin{eqnarray}
G^{-1} & = & g_{\text{S}}^{-1}-\Sigma_{L}-\Sigma_{R}-\Sigma_{\phi}\\
\Sigma_{l;\bs r,\bs r'}\left(z,z'\right) & = & \abs{v_{l}}^{2}\sum_{\alpha}g_{l;\alpha,\alpha}\left(z,z'\right)\delta_{\bs r,\bs r_{l}}\delta_{\bs r',\bs r_{l}}\\
\Sigma_{\phi;\bs r,\bs r'} & = & -\frac{1}{2}\bs{\sigma}.\bs{\phi}_{\bs r}\left(z\right)\delta_{\bs r,\bs r'}\delta\left(z-z'\right)
\end{eqnarray}

\subsection{Properties of the reservoirs \label{sub:Properties-of-the}}

As mentioned in the main text the reservoirs are assumed to be metallic
leads with a constant density of states within all relevant energy
scales. The reservoirs are held in a thermal state characterized by
a chemical potential $\mu_{l}$ and a temperature $T_{l}$. Under
this assumptions we can write 
\begin{eqnarray}
\Sigma_{l}^{R/A}\left(t,t'\right) & \simeq & \mp i\Gamma_{l}\delta\left(t-t'\right)\ket{\bs r_{l}}\bra{\bs r_{l}}\\
\Sigma_{l}^{K}\left(t,t'\right) & \simeq & -2i\Gamma_{l}F_{l}\left(t-t'\right)\ket{\bs r_{l}}\bra{\bs r_{l}}
\end{eqnarray}
with $\Gamma_{l}=\pi\abs{v_{l}}^{2}\rho_{l}\left(0\right)$ and 
\begin{eqnarray}
F_{l}\left(t-t'\right) & = & \int\frac{d\varepsilon}{2\pi}\tanh\left[\frac{\beta_{l}}{2}\left(\varepsilon-\mu_{l}\right)\right]e^{-i\varepsilon t}
\end{eqnarray}

\section{Saddle-Point equations}

\subsection{Variation of the }

We define classical and quantum fields as 
\begin{eqnarray}
\left(\begin{array}{c}
\bs{\phi}_{c,\bs r}^{i}\left(t'\right)\\
\bs{\phi}_{q,\bs r}^{i}\left(t'\right)
\end{array}\right) & = & \frac{1}{\sqrt{2}}\left(\begin{array}{cc}
1 & 1\\
1 & -1
\end{array}\right).\left(\begin{array}{c}
\overrightarrow{\bs{\phi}_{\bs r}^{i}}\left(t'\right)\\
\overleftarrow{\bs{\phi}_{\bs r}^{i}}\left(t'\right)
\end{array}\right)
\end{eqnarray}
where $\overrightarrow{\bs{\phi}_{\bs r}^{i}}\left(t\right),\overleftarrow{\bs{\phi}_{\bs r}^{i}}\left(t\right)=\bs{\phi}_{\bs r}^{i}\left(z\right)$
(for $z\in\gamma_{\rightarrow},\gamma_{\leftarrow}$) are respectively
the Hubbard-Stratonovich fields in the forwards and backwards parts
of the contour. In this way we have that 
\begin{eqnarray}
-\frac{1}{3U}\sum_{\bs r,i}\int_{\gamma}dz\bs{\phi}_{\bs r}^{i}\left(z\right)\bs{\phi}_{\bs r}^{i}\left(z\right) & = & -\frac{1}{3U}\sum_{\bs r,i}\int dt\ \left(\begin{array}{c}
\overrightarrow{\bs{\phi}_{\bs r}^{i}}\left(t\right)\\
\overleftarrow{\bs{\phi}_{\bs r}^{i}}\left(t\right)
\end{array}\right)^{T}\left(\begin{array}{cc}
1 & 0\\
0 & -1
\end{array}\right)\left(\begin{array}{c}
\overrightarrow{\bs{\phi}_{\bs r}^{i}}\left(t'\right)\\
\overleftarrow{\bs{\phi}_{\bs r}^{i}}\left(t'\right)
\end{array}\right)\nonumber \\
 & = & -\frac{1}{3U}\sum_{\bs ri}\int dt\left(\begin{array}{c}
\bs{\phi}_{c,\bs r}^{i}\left(t\right)\\
\bs{\phi}_{q,\bs r}^{i}\left(t\right)
\end{array}\right)^{T}\left(\begin{array}{cc}
0 & 1\\
1 & 0
\end{array}\right)\left(\begin{array}{c}
\bs{\phi}_{c,\bs r}^{i}\left(t'\right)\\
\bs{\phi}_{q,\bs r}^{i}\left(t'\right)
\end{array}\right)
\end{eqnarray}
We proceed to find the saddle-point equations $\delta_{\phi_{a,\bs r}^{i}\left(t\right)}S\left[\phi\right]=0$,
resulting in 
\begin{eqnarray}
\bs{\phi}_{c,\bs r}^{i}\left(t\right) & = & -i\frac{3U}{4}\tr\left\{ \frac{1}{\sqrt{2}}\left[G^{T}\left(t,t+0^{+}\right)+G^{\bar{T}}\left(t+0^{+},t\right)\right]\sigma^{i}\right\} \\
\bs{\phi}_{q,\bs r}^{i}\left(t\right) & = & -i\frac{3U}{4}\tr\left\{ \frac{1}{\sqrt{2}}\left[G^{T}\left(t,t+0^{+}\right)-G^{\bar{T}}\left(t+0^{+},t\right)\right]\sigma^{i}\right\} 
\end{eqnarray}
with $G^{T}$ and $G^{\bar{T}}$ being the propagators on the forward
and backward parts of the contour. Evaluated at the causal solution:
$\bs{\phi}_{q,\bs r}^{i}\left(t\right)=0$ we obtain 
\begin{eqnarray}
\bs{\phi}_{c,\bs r}^{i}\left(t\right) & = & -i\frac{3U}{4}\frac{1}{\sqrt{2}}\tr\left[G^{K}\left(t,t\right)\sigma^{i}\right]
\end{eqnarray}
From these conditions we obtain, at the saddle-point, 

\begin{eqnarray}
\Sigma_{\phi;\bs r,\bs r'}^{R/A}\left(t,t'\right) & = & -\sqrt{2}\delta\left(t-t'\right)\delta_{\bs r,\bs r'}\frac{1}{2}\bs{\sigma}.\bs{\phi}_{c,\bs r}\left(t\right)\\
\Sigma_{\phi}^{K}\left(t,t'\right) & = & 0
\end{eqnarray}

\subsection{Equations of motion \label{sub:Equations-of-motion}}

From Dyson's equation, i.e. $\left[G^{-1}\right]^{R/A}G^{R/A}=1$,
$\left[G^{R}\right]^{-1}G^{K}=\Sigma^{K}G^{A}$ and $G^{K}\left[G^{A}\right]^{-1}=G^{R}\Sigma^{K}$
with $\phi$ evaluated at the saddle-point conditions, we obtain 
\begin{eqnarray}
G^{R}\left(t,t'\right) & = & -i\,\Theta\left(t-t'\right)U\left(t,t'\right)\\
G^{A}\left(t,t'\right) & = & i\,\Theta\left(t'-t\right)\tilde{U}\left(t,t'\right)\\
G^{K}\left(t,t'\right) & = & U\left(t,0\right)G^{K}\left(0,0\right)\tilde{U}\left(0,t'\right)+\int_{0}^{t}d\tau\int_{0}^{t'}d\tau'\, U\left(t,\tau\right)\Sigma^{K}\left(\tau,\tau'\right)\tilde{U}\left(\tau',t'\right)
\end{eqnarray}
where 
\begin{eqnarray}
U\left(t,t'\right) & = & \mathcal{T}e^{-i\int_{t'}^{t}d\tau\, K\left(\tau\right)}\\
\tilde{U}\left(t,t'\right) & = & \left[U\left(t',t\right)\right]^{\dagger}=\mathcal{\tilde{T}}e^{i\int_{t'}^{t}d\tau\, K^{\dagger}\left(\tau\right)}
\end{eqnarray}
are the time order $\mathcal{T}$ and anti-time ordered $\mathcal{\tilde{T}}$
products and

\begin{eqnarray}
\bs K\left(t\right) & = & \bs H\left(t\right)-i\bs{\Gamma}
\end{eqnarray}
with
\begin{eqnarray}
\bs H\left(t\right) & = & \sum_{\bs r\bs r'\sigma}\ket{\bs r,s}\left[-\tilde{t}_{\bs r,\bs r'}-\frac{1}{\sqrt{2}}\delta_{rr'}\bs{\sigma}_{ss'}.\bs{\phi}_{c,\bs r}\left(t\right)\right]\bra{\bs r',s'}\\
\bs{\Gamma} & = & \bs{\Gamma}_{L}+\bs{\Gamma}_{R}\\
\bs{\Gamma}_{l} & = & \Gamma_{l}\ket{\bs r_{l}}\bra{\bs r_{l}}
\end{eqnarray}
is a single-particle operator. With this notation, the many-body operator
$K$ defined in the main text is given by 
\begin{eqnarray*}
K & = & \sum_{\bs r\bs r'ss'}c_{\bs rs}^{\dagger}\bra{\bs rs}\bs K\ket{\bs r's'}c_{\bs r's'}.
\end{eqnarray*}
The equation for $G^{K}\left(t,t\right)$, together with the saddle-point
conditions constitute a closed set that can be used to describe the
evolution of the system at mean-field level: 
\begin{eqnarray*}
\bs{\phi}_{c,\bs r}^{i}\left(t\right) & = & -i\frac{3U}{4}\frac{1}{\sqrt{2}}\tr\left[G^{K}\left(t,t\right)\sigma^{i}\right]\\
G^{K}\left(t,t\right) & = & U\left(t,0\right)G^{K}\left(0,0\right)\tilde{U}\left(0,t\right)-2\pi\int_{0}^{t}d\tau\int_{0}^{t}d\tau'\sum_{l}P\left[\frac{1}{\left(\tau-\tau'\right)}\right]\frac{e^{-i\mu_{l}\left(\tau-\tau'\right)}\frac{\pi\left(\tau-\tau'\right)}{\beta_{l}}}{\sinh\left[\frac{\pi\left(\tau-\tau'\right)}{\beta_{l}}\right]}U\left(t,\tau\right)\bs{\Gamma}_{l}\tilde{U}\left(\tau',t\right)
\end{eqnarray*}
where we used $\int\frac{d\varepsilon}{2\pi}\,\tanh\left[\frac{\beta_{l}}{2}\left(\varepsilon-\mu_{l}\right)\right]e^{-i\varepsilon t}=e^{-i\mu_{l}t}\lim_{\eta\to0}-i\frac{t/\pi}{\left(\eta^{2}+t^{2}\right)}\frac{\frac{\pi t}{\beta_{l}}}{\sinh\left(\frac{\pi t}{\beta_{l}}\right)}=-i\pi e^{-i\mu_{l}t}P\left(\frac{1}{t}\right)\frac{\left(\frac{\pi t}{\beta_{l}}\right)}{\sinh\left(\frac{\pi t}{\beta_{l}}\right)}$.

\subsection{Steady-state \label{sub:Steady-state}}

In a steady-state $\bs{\phi}_{c,r}\left(t\right)=\bs{\phi}_{c,r}$.
Assuming that $\bs K$ is diagonalizable with right and left eigenvectors
\begin{eqnarray}
\bs K\ket{\alpha} & = & \lambda_{\alpha}\ket{\alpha}\\
\bra{\tilde{\alpha}}\bs K & = & \lambda_{\alpha}\bra{\tilde{\alpha}}
\end{eqnarray}
such that $\im\lambda_{\alpha}<0$, we can express it as
\begin{eqnarray}
\bs K & = & \sum_{\alpha}\ket{\alpha}\lambda_{\alpha}\bra{\tilde{\alpha}}
\end{eqnarray}
with the identities 
\begin{eqnarray}
\sum_{\alpha}\ket{\alpha}\bra{\tilde{\alpha}} & = & \sum_{\alpha}\ket{\tilde{\alpha}}\bra{\alpha}=1\\
\braket{\alpha}{\tilde{\alpha}'} & = & \delta_{\alpha\alpha'}
\end{eqnarray}
In this basis we also obtain 
\begin{eqnarray}
G^{R}\left(\omega\right) & = & \left(\omega-\bs K\right)^{-1}=\sum_{\alpha}\ket{\alpha}\left(\omega-\lambda_{\alpha}\right)^{-1}\bra{\tilde{\alpha}}\\
G^{A}\left(\omega\right) & = & \left(\omega-\bs K^{\dagger}\right)^{-1}=\sum_{\alpha}\ket{\tilde{\alpha}}\left(\omega-\bar{\lambda}_{\alpha}\right)^{-1}\bra{\alpha}
\end{eqnarray}
and thus 

\begin{eqnarray}
G^{K}\left(\omega\right) & = & G^{R}\left(\omega\right)F\left(\omega\right)-F\left(\omega\right)G^{A}\left(\omega\right)
\end{eqnarray}
with 
\begin{eqnarray*}
F\left(\omega\right) & = & \sum_{\alpha\alpha'}\ket{\alpha}\frac{-2i\sum_{l}\tanh\left[\frac{\beta_{l}}{2}\left(\omega-\mu_{l}\right)\right]\bra{\tilde{\alpha}}\bs{\Gamma}_{l}\ket{\tilde{\alpha}'}}{\lambda_{\alpha}-\bar{\lambda}_{\alpha'}}\bra{\alpha'}
\end{eqnarray*}

\section{Quadratic approximation to the action around $\phi=0$ \label{sec:quadratic-approximation-to}}

\subsection{Second order contribution }

The second order approximation of the action around $\phi\simeq0$
is given by 

\begin{eqnarray}
S\left[\phi\right] & \simeq & \frac{1}{2}\left[\phi\pi^{-1}\phi\right]-i\,\left\{ \tr\ln\left[-i\left(G_{0}^{-1}\right)\right]-\frac{1}{2}\tr\left[\left(G_{0}\Sigma\right)^{2}\right]\right\} \\
 & = & -i\tr\ln\left[-i\left(G_{0}^{-1}\right)\right]+\frac{1}{2}\sum_{\bs r\bs r'}\int\frac{d\omega}{2\pi}\ \left(\begin{array}{c}
\bs{\phi}_{c,\bs r}^{i}\left(t\right)\\
\bs{\phi}_{q,\bs r}^{i}\left(t\right)
\end{array}\right)^{T}\left(\begin{array}{cc}
0 & \left[\chi^{-1}\right]_{i,j\bs r\bs r'}^{A}\left(t,t'\right)\\
\left[\chi^{-1}\right]_{i,j\bs r\bs r'}^{R}\left(t,t'\right) & \left[\chi^{-1}\right]_{i,j\bs r\bs r'}^{K}\left(t,t'\right)
\end{array}\right)\left(\begin{array}{c}
\bs{\phi}_{c,\bs r}^{i}\left(t'\right)\\
\bs{\phi}_{q,\bs r}^{i}\left(t'\right)
\end{array}\right)
\end{eqnarray}
with $G_{0}^{-1}=\left.G^{-1}\right|_{\phi=0}$. The magnetic susceptibility
is defined as $\chi_{\bs r\bs r'}^{ij}\left(z,z'\right)=-i\av{T_{\gamma}S_{\bs r}^{i}\left(z\right)S_{\bs{r'}}^{j}\left(z'\right)}$.
Explicitly we have 

. 
\begin{eqnarray*}
\left[\chi^{-1}\right]_{\bs r\bs r'}^{ij}\left(t,t'\right) & = & \delta_{ij}\left(\begin{array}{cc}
0 & -\frac{2}{3U}\delta_{\bs r\bs r'}\delta\left(t-t'\right)-\Xi_{ij;\bs r\bs r'}^{A}\left(t,t'\right)\\
-\frac{2}{3U}\delta_{\bs r\bs r'}\delta\left(t-t'\right)-\Xi_{ij;\bs r\bs r'}^{R}\left(t,t'\right) & -\Xi_{ij;\bs r\bs r'}^{K}\left(t,t'\right)
\end{array}\right)
\end{eqnarray*}
where $\Xi$ denotes the bubble-like diagrams 
\begin{eqnarray*}
\Xi_{\bs r\bs r'}^{R}\left(t,t'\right) & = & -i\frac{1}{2}\tr\left[G_{0;\bs r'\bs r}^{A}\left(t',t\right)G_{0;\bs r\bs r'}^{K}\left(t,t'\right)+G_{0;\bs r'\bs r}^{K}\left(t',t\right)G_{0;\bs r\bs r'}^{R}\left(t,t'\right)\right]\\
\Xi_{rr'}^{A}\left(t,t'\right) & = & -i\frac{1}{2}\tr\left[G_{0;\bs r'\bs r}^{R}\left(t',t\right)G_{0;\bs r\bs r'}^{K}\left(t,t'\right)+G_{0;\bs r'\bs r}^{K}\left(r't',rt\right)G_{0;\bs r\bs r'}^{A}\left(t,t'\right)\right]\\
\Xi_{rr'}^{K}\left(t,t'\right) & = & -i\frac{1}{2}\tr\left[G_{0;\bs r'\bs r}^{A}\left(t',t\right)G_{0;\bs r\bs r'}^{R}\left(t,t'\right)+G_{0;\bs r'\bs r}^{R}\left(t',t\right)G_{0;\bs r\bs r'}^{A}\left(t,t'\right)+G_{0;\bs r'\bs r}^{K}\left(t',t\right)G_{0;\bs r\bs r'}^{K}\left(t,t'\right)\right]
\end{eqnarray*}
Assuming a steady state condition we obtain, for the retarded component
\begin{eqnarray*}
\Xi_{rr'}^{R}\left(\omega\right) & = & \Xi_{rr'}^{(1)}\left(\omega\right)+\bar{\Xi}_{rr'}^{(2)}\left(-\omega\right)+\Xi_{rr'}^{(2)}\left(\omega\right)+\bar{\Xi}_{rr'}^{(1)}\left(-\omega\right)\\
\Xi_{rr'}^{(1)}\left(\omega\right) & = & -\sum_{\alpha\beta}\sum_{l}\braket{r'}{\tilde{\beta}}\braket{\beta}r\braket r{\alpha}A_{\alpha r'}^{l}\, I_{l}\left(\bar{\lambda}_{\beta}+\omega,\lambda_{\alpha}\right)\\
\Xi_{rr'}^{(2)}\left(\omega\right) & = & -\sum_{\alpha\beta}\sum_{l}\braket{r'}{\alpha}\braket r{\beta}\braket{\tilde{\beta}}{r'}A_{\alpha r}^{l}\, I_{l}\left(\lambda_{\beta}-\omega,\lambda_{\alpha}\right)
\end{eqnarray*}
with 
\begin{eqnarray*}
I_{l}\left(z,z'\right) & = & \frac{1}{\pi}\frac{\psi^{(0)}\left[\frac{1}{2}-i\text{sgn}\left(\im z'\right)\frac{\beta_{l}(z'-\mu_{l})}{2\pi}\right]-\psi^{(0)}\left[\frac{1}{2}-i\text{sgn}\left(\im z\right)\frac{\beta_{l}(z-\mu_{l})}{2\pi}\right]}{z-y}\\
A_{\alpha r}^{l} & = & \sum_{\alpha'}\frac{\bra{\tilde{\alpha}}\bs{\Gamma}_{l}\ket{\tilde{\alpha}'}\braket{\alpha'}r}{\lambda_{\alpha}-\bar{\lambda}_{\alpha'}}
\end{eqnarray*}
with $\psi^{(0)}\left(z\right)=\pd_{z}\ln\Gamma\left(z\right)$ being
the logarithmic derivative of the Gamma function. 

\end{widetext}
\end{document}